# From Code Smells to Best Practices: Tackling Resource Leaks in PyTorch, TensorFlow, and Keras

Bashar Abdallah[1], Martyna E. Wojciechowska[1], Gustavo Santos[1], Edmand Yu[1], Maxime Lamothe[1], Alain Abran[2], Mohammad Hamdaqa[1]

**Abstract** – Much of the existing Machine Learning (ML) research focuses on ML performance and evaluation metrics, such as recall, accuracy, and precision; there is a discernible gap in addressing the long-term sustainability, efficiency, and overall quality of ML applications. Although high performance is essential, it is equally important to ensure that ML applications are resource-efficient, sustainable, and robust for successful deployment in real-world applications. Implementation of software engineering best practices, including effective resource management and sustainability, is crucial for developing mature ML solutions.

This study addresses this gap by systematically identifying bad coding practices (code smells) that contribute to resource leaks, which can significantly impact the efficiency of ML applications. An empirical investigation was conducted on the applications developed using PyTorch, TensorFlow, and Keras. The analysis identified 30 bad coding practices associated with PyTorch and 16 with TensorFlow and Keras.

The study categorized the identified code smells in two distinct ways. The first classification pertains to the root causes of resource leaks, whereas the second is based on general ML smells that are specific to a framework. Furthermore, this study investigated best practices for preventing resource leakage. For each identified code smell, at least one corresponding best practice was recommended, resulting in a total of 50 coding patterns designed to enhance resource efficiency. To ensure the reliability and validity of the findings, a comprehensive three-phase validation process was conducted involving three authors who independently analyzed the data systematically, followed by discussions to reach a consensus.

This study makes a significant contribution to the field of ML by being the first to comprehensively investigate code smells that lead to resource leaks in ML applications, particularly those developed using PyTorch, TensorFlow, and Keras. The proposed best practices offer practical guidance to developers, thereby enhancing the efficiency and sustainability of ML applications. In Additionally, this study provides two categorization styles to elucidate the root causes of resource leaks in ML applications.

Corresponding Author: Bashar Abdallah: bashar.abdallah@etud.polymtl.ca
Martyna E. Wojciechowska: martyna-ewa.wojciechowska@polymtl.ca
Gustavo Santos: gustavo.palma-dos-santos@etud.polymtl.ca
Edmand Yu: edmand.yu@polymtl.ca
Maxime Lamothe: maxime.lamothe@polymtl.ca
Alain Abran: alain.abran@etsmtl.ca
Mohammad Hamdaqa: mhamdaqa@polymtl.ca

[1]**Polytechnique Montréal, Montréal, Canada**
[2]**Ecole de Technologie Superieure – ETS, Montréal, Canada**

# 1 Introduction

The role of software in modern society has expanded rapidly, with governments, industries, and individuals increasingly relying on software applications to deliver critical services. Among these, artificial intelligence (AI) systems, particularly those powered by machine learning (ML), have experienced growing adoption across various domains, including healthcare, education, finance, and security. These ML applications often involve computationally intensive tasks such as image processing, natural language processing, and real-time video analysis, all of which require substantial hardware resources, including GPUs and cloud-based infrastructure.

The heavy resource demands of ML applications make them particularly susceptible to performance degradation when system resources are not managed effectively (Ghanavati et al. 2020). Resource situations where allocated resources, such as memory, file handles, or GPU sessions, are not properly released can lead to decreased system performance, increased operational costs, and even application crashes. Microsoft has identified leak detection and localization as one of the top ten challenges in software development (LO ET AL. 2015).

Code smells are structures that violate fundamental design principles and have a negative impact on design quality (Suryanarayana et al. 2014). In other words, code smells, which are design issues in code that may not cause immediate bugs but can lead to long-term problems such as resource leaks, have been identified as a major contributor to poor software quality. These smells are often due to deviations from best practices in resource allocation and disposal. While code smells are widely recognized in software engineering, their impact on ML applications remains underexplored. This paper investigates common code smells in ML applications that lead to resource leaks, categorizes them, and proposes best practices for mitigation.

The unique characteristics of ML applications—including high memory consumption, extended execution time, and real-time processing requirements—heighten the risks associated with resource leaks. For instance, in reinforcement learning, prolonged training and iterative updates can accumulate leaks over time, compromising learning stability and resource efficiency. In real-time systems, even minor leaks can violate latency constraints, affecting application reliability.

This study presents an empirical investigation aimed at identifying code smells that lead to resource leaks in ML applications. By analyzing discussions on Q&A platforms related to three widely used ML frameworks—PyTorch, TensorFlow, and Keras—we systematically identify recurrent code smells that cause resource leaks and categorize them. Next, we propose the best practices tailored to each discovered smell to mitigate these issues.

These three ML frameworks were selected due to their prominence and broad adoption in both research and industry:

- PyTorch is known for its dynamic computational graph and extensive GPU usage, making it a popular choice for rapid prototyping and experimentation.
- TensorFlow, developed by Google, is widely adopted in production systems and supports both low- and high-level programming models.
- Keras serves as a high-level API built on TensorFlow, offering an accessible interface for building and training deep learning (DL) models.

The empirical study reported here was conducted to uncover the issues that lead to resource leaks. Moreover, it investigates which bad coding practices (code smells) lead to resource leaks and which best practices (patterns) help avoid them. This empirical study aimed to answer the following research questions:

- **RQ1:** What are the specific code smells that contribute to resource leak incidents in PyTorch, TensorFlow, and Keras-based ML applications?
  **Motivation:** ML applications are often developed by practitioners with limited formal training in software engineering (Simmons et al. 2020), increasing the risk of introducing code smells that lead to resource leaks. Identifying these specific smells is a necessary first step toward understanding common pitfalls in ML resource management and minimizing resource leaks in widely used frameworks.
- **RQ2:** Under which category can the identified code smells be categorized?
  **Motivation:**
  Categorization provides a structured understanding of resource leak-related code smells, enabling developers to recognize recurring patterns and apply targeted mitigation strategies to address them. Without categorization, such issues may be inconsistently addressed or overlooked entirely. This study introduces a dual categorization approach: (1) root cause categories based on analyzing the root cause of resource leak cases, and grouping similar causes into one category, and (2) framework specificity—distinguishing whether a smell is common across ML applications or specific to a particular framework (PyTorch, TensorFlow, or Keras). This additional categorization helps distinguish whether resource leakage issues stem from developer practices or limitations within ML frameworks.
- **RQ3:** What are the recommended best practices (coding patterns) for preventing code smells associated with resource leaks in PyTorch, TensorFlow, and Keras-based ML applications?
  **Motivation:** Addressing resource leaks requires actionable, framework-specific guidance. This question aims to identify and synthesize the best practices that developers can adopt to address leaks proactively.

Providing these practices contributes to the development of more sustainable and efficient ML applications and promotes better software engineering practices within the ML community.

## 2 Related Work

In the software engineering field, the term "code smell" was initially introduced to describe poor designs in source code (Fowler 1997). In the literature, researchers have explored the consequences of code smells in software systems, including maintainability (Deligiannis et al. 2004)(Li and Shatnawi 2007)(Yamashita and Moonen 2012). They have also delved into the timing and reasons behind the introduction of code smells (Tufano et al. 2015) and various techniques for detecting code smells (Munro 2005; Lanza and Marinescu 2006; Moha et al. 2010). Researchers have increasingly turned to ML models as powerful tools for detecting memory leaks in software. For example, (Andrzejak et al. 2017) utilized ML techniques to identify memory leaks in C/C++ programs. By collecting data on how memory is allocated and released, they trained an ML model to identify the differences between problematic and normal memory usage.

This approach highlights the potential of ML models to improve software reliability and performance by addressing memory leak issues. Also, ML algorithms are used to detect resource leaks in cloud-based infrastructures. For example,(Jindal et al. 2021) introduced two novel algorithms, Linear Backward Regression (LBR) and Precog, along with their variants, to identify memory leaks using only the system's memory utilization data. The approach demonstrated high accuracy and efficiency. The PrecogMF algorithm achieved an 85% accuracy rate and significantly reduced compute time. This study highlights the potential of ML models to enhance the reliability and performance of cloud-based applications by effectively addressing memory leak issues.

The literature review includes an investigation of the ML frameworks. (Ajel et al. 2023) performed a comprehensive analysis of four popular frameworks: TensorFlow, Keras, PyTorch, and Scikit-learn. That study evaluates these frameworks based on key metrics such as energy efficiency, memory usage, execution time, and accuracy. The findings revealed significant trade-offs, with slower frameworks generally consuming less energy and faster frameworks using more memory. Specifically, PyTorch demonstrated balanced performance with moderate energy consumption and memory usage, making it a suitable choice for energy-aware applications. In contrast, TensorFlow showed higher energy consumption but offered faster execution times. Keras and Scikit-learn exhibited varying performance depending on the specific tasks and datasets used, underscoring the importance of considering energy efficiency in developing and deploying ML models and promoting energy-aware programming practices. The insights from that study contribute to a broader understanding of software sustainability in ML applications, particularly in PyTorch. However, code smells and resource leaks were not covered in that study.

In addition, (Sun et al. 2021) investigated ML libraries, including TensorFlow, PyTorch, and Scikit-learn, to examine bug categories, fixing patterns, and maintenance types in ML projects. However, comprehensive reviews that delve into how code smells specifically contribute to resource leaks within these frameworks are scarce. Addressing this gap could provide valuable insights into improving the reliability and efficiency of ML applications.

Code smells have been investigated in ML applications. For instance, the analysis in (Van Oort et al. 2021) of 74 open-source ML projects using Pylint to identify code smells, potential defects, refactoring opportunities, and coding standard violations in ML projects reported a list of the top 20 detected code smells per category, and common issues such as code duplication and challenges in applying the PEP8 convention to ML code.

The study in (Tang et al. 2021) investigated the gap in understanding how ML systems evolve and are maintained by studying real-world open-source software and the refactorings used to address technical debt. The analysis involved 26 projects with 4.2 million lines of code and 327 manually examined code patches. The study reported that developers perform refactorings for various reasons, some of which relate to technical debt, while others do not. Code duplication, especially in ML configuration and model code, is a common issue and a major focus of refactoring efforts. Their primary focus was on the classification of distinct refactoring types. However, the study did not extract the specific code smell or patterns that should prompt the initiation of such a refactoring process.

The empirical study in (Zhang et al. 2022) identified code smells specific to ML applications. A comprehensive dataset of code smells was collected from various sources, including academic papers, grey literature entries, bug reports from Stack Overflow and GitHub, and additional related Stack Overflow discussions. The study goal was to initiate a dialogue regarding ML-specific code smells and contribute to enhancing code quality within the ML community. A total of 22 code smells, such as unnecessary iterations, encompassing both general and API-specific issues, were identified and categorized according to different pipeline stages and their respective impact.

The study in (Gesi et al. 2022) investigated code smells as potential indicators of issues in ML systems. It identified nine common maintenance-related changes made by ML developers and summarized five ML-specific code smells, such as Scattered Use of ML Library, Unwanted Debugging Code, etc. These findings validated the prevalence and impact of these code smells, demonstrating that they have a significant impact on ML system maintenance from a developer's standpoint.

The literature review of related work highlights a conspicuous lack of exploration of resource leak issues within ML applications. None of the scrutinized studies have addressed the specific challenges and intricacies related to resource leaks in ML applications, encompassing their code smells and patterns. This underscores a notable gap in the existing knowledge base and the need to investigate resource leaks in ML applications. Moreover, research conducted on ML libraries shows a conspicuous lack of investigation into code smells that lead to resource leaks. To the best of our knowledge, this research study, as reported here, represents the first exploration of code smells that contribute to resource leaks in ML applications.

## 3 Research Methodology

This study analyzes ML developers' discussions on community-driven platforms, including the PyTorch Q&A platform, the TensorFlow Q&A platform, and Stack Overflow (for both TensorFlow and Keras), which were selected for their popularity and active developer participation. Stack Overflow was selected as the primary source for Keras-related discussions due to the absence of a standalone official forum dedicated exclusively to Keras. For TensorFlow, Stack Overflow posts were also included because the official TensorFlow forum contained limited discussion on resource leak issues.

A Python script was used to extract posts related to resource and memory leaks by searching for the keywords 'resource leaks,' 'memory leaks,' and 'out of memory' on the selected platforms. These keywords were chosen because developers commonly use them to describe resource leak issues in ML applications. This process yielded 671 posts from the PyTorch platform, 219 posts from both the TensorFlow platform and Stack Overflow, and 81 Keras-related posts from Stack Overflow. The analysis focuses on real-world code snippets and developer-reported issues to detect common patterns and root causes of resource leaks. An overview of the research methodology is presented in Figure 1.

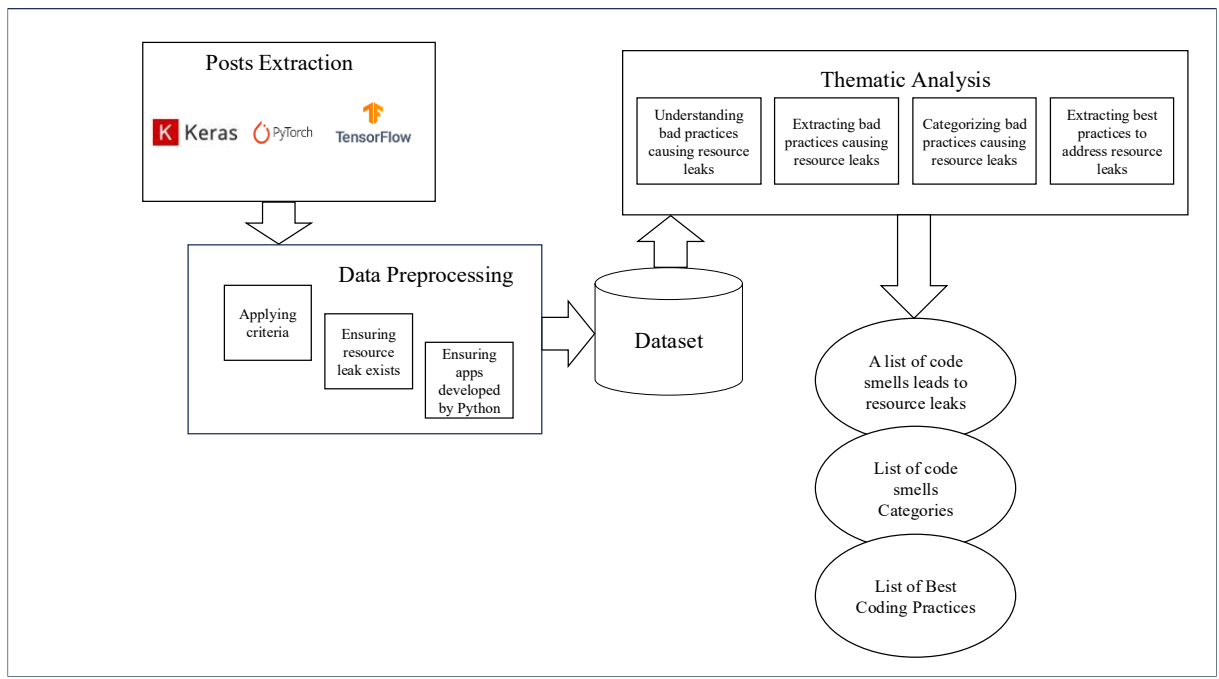

Figure 1: Research methodology

Thematic analysis was employed to identify code smells and corresponding best practices. Keywords identified during the coding phase guided the development of themes, which were iteratively refined and categorized based on the root causes of the resource leaks. For instance, initial categories based on syntactic patterns (e.g., loop-related keywords) were reclassified according to their underlying causes after deeper analysis. To support the identification of patterns, color coding was applied, with each theme assigned a distinct color to represent recurring practices and their implications visually. This method facilitated the recognition of problematic coding patterns and supported the extraction of consistent themes across different frameworks. Manual validation was conducted at multiple stages to ensure the relevance, quality, and framework consistency of the selected posts. Figure 2 provides an example of analyzing a post.

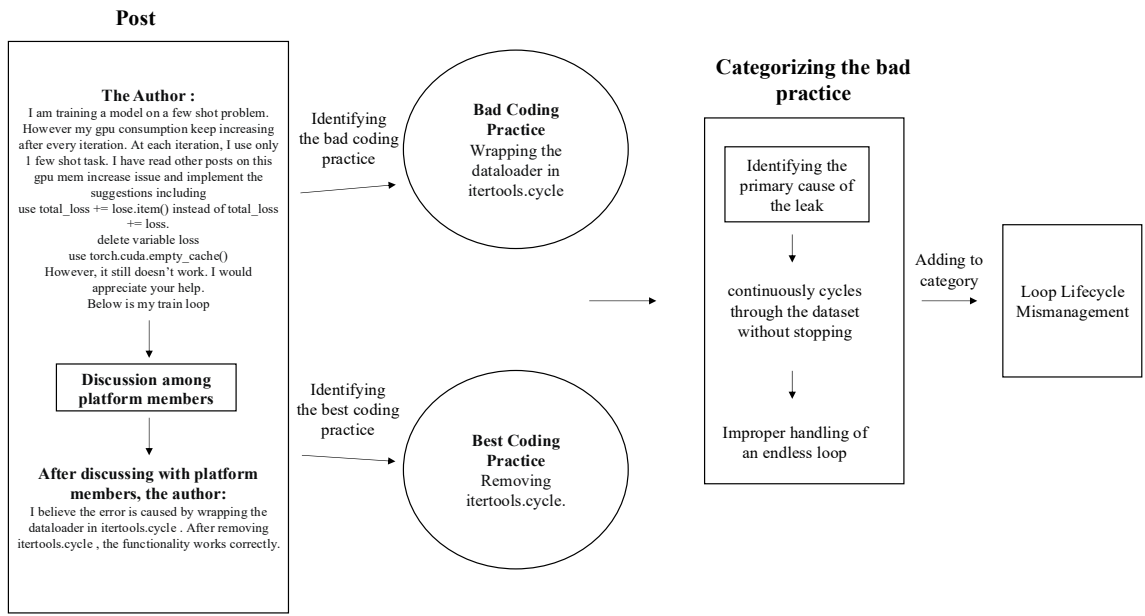

Figure 2: Example of post-analysis

Not all posts are clearly explained as shown in Figure 2. Some authors refer to best practices found in other posts or link to external sources. In such cases, additional effort was required by the researchers to trace the root cause of the leak and understand the recommended solution by reviewing the referenced materials. In more complex scenarios, the categorization process was not as straightforward as the example in Figure 2, as some issues were associated with multiple root causes. Further discussion of these complexities is provided in the discussion (section 5).

To ensure the quality and relevance of the data, specific inclusion criteria were applied. Since no posts related to PyTorch existed before 2017, we used this criterion for all frameworks to ensure they were evaluated within the same time frame. The selected posts in which developers reported experiencing resource or memory leaks and detailed their resolution methods are of particular interest because these discussions provide valuable insights into potential code smells and effective mitigation strategies. The dataset spans posts published between 2017 and 2024 across all three frameworks. Manual validation was conducted to verify that each post met the inclusion criteria. Table 1 summarizes the selection criteria.

Table 1: Selection criteria

| No. | Criteria | Comments |
|---|---|---|
| 1. | The leak is solved | Posts that still have unresolved resource leak issues are not considered in this study. |
| 2. | Published between January 2017 and December 2024 | |
| 3. | Developed in Python | |
| 4. | Resource leaks exist | |

Upon further analysis, it was notice that many posts were submitted by users who believed their applications suffered from resource leaks; however, responses often revealed that the underlying issues were related to performance optimization rather than actual leaks. In PyTorch, many posts were excluded due to confusion surrounding the term "leak" in the context of the LeakyReLU activation function. Although these posts matched the keyword filter, they did not pertain to resource leaks and were therefore removed from the dataset. As a result, the final dataset included 65 relevant posts for PyTorch, 27 for TensorFlow, and 10 for Keras.

To enhance the validity of the study and reduce potential bias, three authors were involved in the analysis process. Each step of the research methodology was independently applied by the three authors. The results were then compared and validated through discussion to resolve discrepancies. This collaborative approach was adopted to ensure greater objectivity and reliability in the findings.

In several cases, posts were excluded due to duplication, particularly within the Keras and TensorFlow categories, where overlap is common due to Keras's integration with TensorFlow. To accurately identify duplicates, each post was tagged with a unique identifier corresponding to its platform-specific post ID. This ensured that no post appeared more than once in the dataset. This deduplication process accounts for a significant portion of the difference between the number of posts initially collected and those ultimately included in the analysis.

To classify a case as a resource leak, several criteria were applied. A resource leak was defined as a scenario in which an application continuously consumes memory or other resources without proper release, ultimately leading to resource exhaustion. Specifically, cases were identified as resource leaks when resources such as memory, GPU memory, file handles, or network connections were not correctly deallocated, causing the application to retain unused resources (Shadab et al. 2023). Additionally, instances where performance degradation or system crashes occurred due to the exhaustion of critical resources, such as memory or file descriptors, were also classified as resource leaks (Shadab et al. 2023). Supporting indicators included memory fragmentation, error messages related to resource exhaustion, and persistent resource usage identified through performance profiling. These criteria enabled clear differentiation between genuine resource leaks and other issues, such as optimization inefficiencies.

# 4. Results and Findings

This section provides answers to the research questions, with each response based on the findings from the three frameworks studied.

- RQ1: What are the specific code smells that contribute to resource leak incidents in PyTorch, TensorFlow, and Keras-based ML applications?

To address RQ1, this study identifies specific code smells that contribute to resource leaks in ML applications. By systematically analyzing real-world cases, we identify recurring patterns of poor coding practices that result in inefficient resource management. These patterns reveal common vulnerabilities in ML code and inform the development of best practices.
The findings provide actionable insights to enhance code efficiency and prevent resource leaks in PyTorch, TensorFlow, and Keras-based applications. Figure 3 summarizes the identified code smells contributing to resource leaks in PyTorch-based applications.

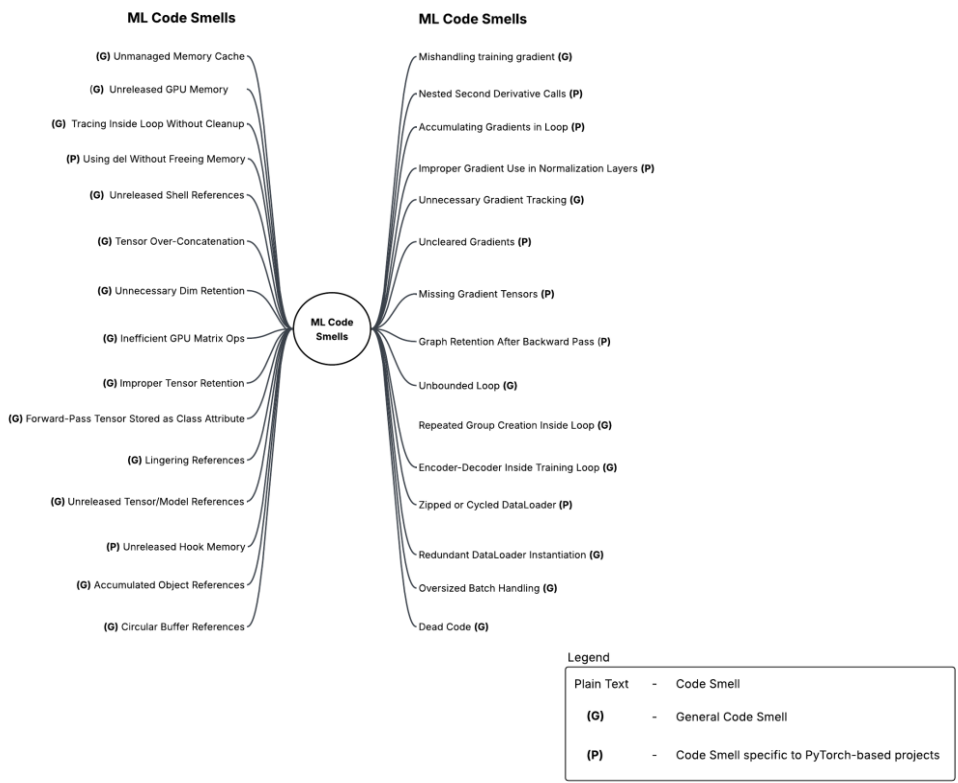

Figure 3: Code smells identified in PyTorch-based applications

Figure 3 provides an overview of the identified code smells in PyTorch-based applications. These code smells indicate potential issues leading to inefficient memory usage, resource retention, and performance degradation:

- **Unreleased GPU Memory:** Running the model without appropriately releasing GPU memory can result in inefficient memory utilization and the onset of memory leaks. Specifically, the `rpn._cache` attribute, which stores pairs with the model, becomes problematic as the number of tensors it retains grows with an increasing number of proposals. As the cache expands, memory is progressively consumed without proper deallocation, leading to a gradual buildup of unused resources. This unchecked accumulation of memory can cause significant performance degradation, elevating memory consumption and increasing the risk of out-of-memory (OOM) errors. Over time, such memory leaks undermine the stability and efficiency of the model's operations, particularly when processing large datasets or performing extended inference tasks.
- **Graph Retention After Backward Pass:** Retaining the computational graph after the completion of a

backward pass, when not required, can result in substantial memory inefficiencies and potential resource leaks in PyTorch-based models. The computational graph holds intermediate variables and gradients necessary for backpropagation, and preserving it unnecessarily leads to continued memory allocation that is not reclaimed. This issue becomes especially pronounced in contexts involving large datasets or deep neural networks, where the size and complexity of the graph can lead to excessive memory consumption. Over time, this practice may cause memory to be retained beyond its useful scope, resulting in resource leaks that degrade performance and, in severe cases, lead to out-of-memory failures.

- **Unbounded Loop:** Improper handling of an endless loop can arise when functions like `itertools.cycle` are employed, resulting in an infinite loop over the data loader.
  This practice leads to unnecessary consumption of resources and hinders the system's ability to process the data efficiently.
- **Accumulating Gradients in Loop:** Computing gradients in a loop without resetting gradients or detaching tensors results in inefficient resource management, as gradients from earlier iterations accumulate, consuming additional memory and degrading system performance.
- **Unnecessary Gradient Tracking:** Excessive reliance on gradient tracking management (`torch.no_grad()`) in PyTorch, particularly when it is either not applied where necessary or used inconsistently across the code, can lead to unnecessary resource retention, including excessive memory consumption and computational overhead. Specifically, when gradients are tracked during inference or evaluation, where they are not required, this results in inefficient resource utilization. Operations that do not necessitate backpropagation continue to have gradients computed and stored, thereby degrading performance and increasing memory usage.
- **Release Memory Failure**: Failure to release memory after inference operations can result in resource leaks, as memory consumption increases without proper deallocation. Although the model is no longer trained, intermediate tensors and computation graphs may persist in memory, thereby occupying valuable resources. This continued allocation leads to gradual memory buildup, which exacerbates the risk of out-of-memory (OOM) errors, particularly when inference is conducted on large datasets or over extended durations. As inference operations accumulate, the accumulation of unused memory further degrades system performance and may ultimately lead to crashes or significant slowdowns.
- **Unreleased Hook Memory:** Forward hooks are employed to capture the inputs or outputs of a module during the forward pass; however, if the memory utilized by these hooks is not properly deallocated after their use, it can lead to the gradual accumulation of GPU memory. Each invocation of a forward hook may result in additional memory allocations, and in the absence of proper resource deallocation, the memory footprint unnecessarily expands. This accumulation of unused memory can impair system performance, heighten the risk of OOM errors, and, in extreme cases, lead to the complete exhaustion of GPU memory, thus compromising the stability and efficiency of both model training and inference processes.
- **Redundant DataLoader Instantiation:** Improper management of DataLoader with multiple workers (e.g., `persistent_workers=True`) can result in inefficient resource utilization, leading to unnecessary memory overhead, synchronization bottlenecks, or excessive consumption of CPU/GPU resources. When the number of workers is not optimally configured to match the system's hardware capabilities, it may cause thread contention, where workers compete for limited resources, thereby decreasing overall throughput. Furthermore, failure to properly terminate workers or release allocated memory can result in memory leaks, contributing to performance degradation. These inefficiencies can substantially delay the data loading process, undermining the effectiveness of parallel processing and ultimately impeding the efficiency of model training or inference tasks.
- **Unreleased Tensor/Model References**: Improper handling of tensors and resources can give rise to several detrimental practices that result in inefficiencies and potential resource leaks. For instance, failing to detach tensors from the computation graph leads to the persistence of the graph in memory, preventing the release of unused resources and causing excessive memory consumption. Similarly, not removing models or computation graphs after use can result in retained references, contributing further to memory buildup. Inefficient tensor transfers between the GPU and CPU also degrade performance, as frequent data movement introduces significant overhead that slows computational processes. Moreover, inadequate management of circular dependencies or gradients during backpropagation can inadvertently lead to memory retention, obstructing proper memory deallocation and exacerbating resource leaks. These practices impede performance, elevate memory usage, and can undermine the stability and efficiency of ML workflows.
- **Zipped or Cycled DataLoader:** Zipping or cycling an image DataLoader can result in inefficient resource utilization and potential data loading issues by introducing unnecessary repetition of data loading or inefficient iteration patterns. The zip() function combines multiple iterators, such as different DataLoader objects, into a single iterable. This can lead to inefficiencies if the data from different sources does not align correctly, causing mismatched batch sizes or improper synchronization, which hinders efficient parallel processing. In contrast, the cycle() function creates an infinite iterator, repeatedly cycling through the dataset. While this may be useful in certain contexts, cycling a DataLoader can cause the continuous loading of identical data without deallocating resources, thereby increasing memory consumption and computational overhead. Such practices can introduce significant inefficiencies and performance bottlenecks, particularly when working with large datasets or complex models.

- **Tensor Over-Concatenation:** The overuse or frequent concatenation of tensors along a specified dimension, particularly through the repeated application of `torch.cat`, can result in substantial performance inefficiencies and increased memory consumption. Each concatenation operation entails the creation of a new tensor, which necessitates the allocation of additional memory and the copying of data from the original tensors. When such operations are performed repeatedly, they not only increase the memory footprint but also impose computational overhead due to continuous memory reallocations. As the tensors grow in size, the concatenation process becomes progressively more costly in terms of both time and memory usage. Furthermore, excessive tensor concatenation can lead to memory fragmentation, complicating the efficient management of resources and potentially causing out-of-memory (OOM) errors. These practices, therefore, detrimentally affect the overall efficiency of model training or inference pipelines, particularly when working with large datasets or complex models.
- **Unreleased Shell References:** Creating and managing variables directly within interactive Python shells (e.g., IPython) is considered a poor coding practice that may lead to resource leaks, particularly in long-running sessions. Unlike scripts where variables are typically scoped and managed within functions or modules, interactive shells lack structured cleanup mechanisms. As a result, variables instantiated during exploratory coding are often left undeleted or persist unintentionally in memory. Over time, this accumulation of unused or obsolete variables can contribute to significant memory consumption, especially when handling large datasets or model components.
- **Improper Tensor Retention:** Directly storing tensors in a data structure without employing proper context management can result in memory leaks. Specifically, saving tensors directly in the `saved_data` hash-map may lead to the accumulation of intermediate results, such as `ctx.x` or `ctx.z`, which are not appropriately released across multiple mini-batches. As these tensors accumulate without proper deallocation, GPU memory is progressively consumed, leading to heightened memory usage. This improper management of tensor storage disrupts efficient memory handling and adversely affects the stability and performance of the model.
- **Dead Code:** leaving debugging-related code, such as `torch.autograd.detect_anomaly(True)`, in production or training code constitutes a poor practice due to the potential for unnecessary overhead and performance degradation. The `torch.autograd.detect_anomaly(True)` function is valuable during debugging for identifying and reporting anomalies in the computation graph, such as NaN or infinite values in gradients. However, if retained in the code after the debugging phase, it introduces redundant checks and computations that are unnecessary for standard training or inference. This results in a notable reduction in model performance, as it continuously monitors gradients for anomalies, thereby consuming additional computational resources.
- **Unnecessary Dim retention:** Unnecessary retention of dimensions occurs when tensor dimensions are preserved following a reduction operation, despite the fact that these dimensions are not required for subsequent computations. This is typically achieved by retaining the reduced dimensions with a size of 1, thereby unnecessarily increasing the tensor's size. While this practice may appear inconsequential, it leads to inefficient memory utilization, as the additional dimensions consume memory without serving a functional purpose. Furthermore, the retention of these superfluous dimensions can result in heightened computational overhead, as operations on larger tensors require more time and resources. This practice can significantly degrade performance if not appropriately managed, particularly in scenarios involving large datasets or frequent reduction operations.
- **Circular Buffer References:** The creation of circular references between buffers and objects constitutes a detrimental practice that inhibits the proper deallocation of memory, ultimately leading to memory leaks. A circular reference arises when two or more objects reference each other in such a manner that they cannot be collected as garbage, as they maintain mutual dependencies. In the context of ML, this issue may occur when buffers, which store intermediate data or model parameters, and objects, such as layers or models, form circular dependencies. Specifically, when one object holds a reference to a buffer while the buffer reciprocates by referencing the object, both remain in memory even when no longer necessary. As a result, memory that should be freed remains allocated, causing a gradual accumulation of unused resources.
- **Accumulated Object References:** Improper handling of accumulated references is a problematic practice that can lead to significant memory management issues. For instance, in custom autograd functions, objects stored using self.save_for_backward() can accumulate over time without being properly cleared or managed. However, this issue is not limited to custom autograd functions. It can occur in various situations where intermediate results or tensors are unnecessarily retained in memory. When references to objects are not cleared after they are no longer needed, memory is not deallocated, causing inefficient memory usage and potential memory leaks. Over time, such accumulation of unnecessary references leads to increased memory leaks.

- **Inefficient GPU Matrix Ops:** performing matrix multiplication operation (e.g., using torch.mm) directly on two 2D tensors on the GPU, without appropriate memory management, exacerbates memory-related issues. This practice can lead to substantial CUDA memory leaks due to the improper handling of intermediate memory allocations during the matrix multiplication operation. When intermediate tensors are not properly released following the operation, GPU memory is not reclaimed, leading to the progressive exhaustion of available resources.
- **Lingering References:** Lingering tensor references occur when PyTorch tensors, no longer needed, continue to be stored or referenced in memory. These persistent references prevent the garbage collector from deallocating the memory they occupy, resulting in unused memory remaining allocated. In the case of Deep Q-Network (DQN) replay memory, this issue can lead to excessive memory consumption. When PyTorch tensors are stored in replay memory without proper management, they retain references that hinder automatic garbage collection, causing memory to accumulate unnecessarily.
- **Forward-Pass Tensor Stored as Class Attribute:** Using a class attribute to hold a tensor reference within the forward pass constitutes a suboptimal memory management practice. By storing a tensor as a class attribute (e.g., self._ten), the tensor reference is retained throughout the object's lifetime, even after the tensor is no longer required for subsequent computations. This persistent reference prevents the automatic deallocation of memory, resulting in unused tensors occupying memory, which increases memory consumption and may lead to memory leaks. As the tensor is unnecessarily retained, system performance may degrade due to inefficient memory utilization, particularly in long-running tasks or large-scale models.
- **Missing Gradient Tensors:** Failing to store tensors needed for gradient computation during the backward pass using ctx.save_for_backward() introduces memory management inefficiencies. This function is designed to track tensors necessary for the backward pass, ensuring proper handling of their deallocation once they are no longer required. Without it, tensors critical for gradient computation may not be properly tracked or released, leading to an accumulation of unreferenced tensors that persist in memory.
- **Unreleased GPU Memory:** The improper handling of the training model's gradient during the inference phase, specifically by not correctly disabling gradient computation, constitutes a detrimental practice in model deployment. During training, gradients are necessary for parameter updates via backpropagation, but during inference, these gradients are not required. The use of torch.no_grad() is critical to disable gradient tracking and prevent unnecessary computation. However, when torch.no_grad() is not properly set to True for the parameters of a pre-trained model during inference, the model continues to track gradients, leading to the retention of the computation graph. This improper handling results in increased memory consumption as unused gradients are stored, thereby degrading system performance. Additionally, the unnecessary computation of gradients introduces inefficiencies that slow down the inference process, increase memory usage, and may lead to potential memory leaks over time. Therefore, it is essential to appropriately manage gradient computation during inference to optimize resource utilization and maintain efficient performance in ML models.
- **Oversized Batch Handling:** Inadequate handling of large tensors, extended input sequences, or substantial batch sizes in DL workflows can lead to significant resource leaks stemming from inefficient memory utilization. DL models—particularly those processing high-dimensional data or performing complex computations—consume considerable GPU or CPU memory. When such large data structures are not explicitly managed, released, or reused appropriately, memory may be retained unnecessarily, even after the associated computations are complete. This retained memory, often a result of lingering references or improper garbage collection, can accumulate over time, leading to degraded performance and, in severe cases, out-of-memory errors. Consequently, failure to manage these resources systematically constitutes a critical source of memory leaks in ML applications.
- **Uncleared Gradients:** In PyTorch, the retain_graph=True argument is used with the backward() function to instruct the framework to preserve the computational graph after the backward pass is executed. This is necessary in models that require multiple backward passes through the same graph, such as certain recurrent neural networks or architectures with multiple loss terms. By default, PyTorch frees the graph after a backward call to reclaim memory. However, when retain_graph=True is used unnecessarily or misused—particularly without proper control over repeated use—the retained graph and all associated intermediate tensors and activations are kept in memory. If these elements are not explicitly cleared or are continuously accumulated across iterations without being used or freed, they persist in memory beyond their intended lifetime. This persistent retention of computational resources constitutes a resource leak, as memory that should be released remains occupied, leading to progressive memory bloat. Over time, especially in long training loops or experiments, this can exhaust available GPU memory and cause the application to crash or degrade in performance. Therefore, unless multiple backward passes are explicitly required, the use of retain_graph=True should be avoided to prevent such unintended resource leaks.

- **Nested Second Derivative Calls:** Computing higher-order derivatives through nested invocations of PyTorch's grad() function can lead to inefficient resource utilization and potential memory leaks. Each call to grad() constructs a new intermediate computational graph to support the derivative computation. When these calls are repeated without proper handling, the resulting graphs may not be released promptly, leading to unnecessary retention of memory. This behavior can significantly inflate memory consumption, particularly in models involving large parameter spaces or complex operations. Furthermore, repeated and unoptimized use of nested grad() calls introduces substantial computational overhead, which not only impacts performance but also increases the risk of memory exhaustion. Consequently, careful management of computational graph retention is essential when performing higher-order differentiation to avoid inadvertent resource leaks.

- **Improper Gradient Use in Normalization Layers:** In normalization layers, variables such as self.running_mean and self.running_covar are typically used to store running estimates of the input's mean and covariance, respectively. These statistics are updated during training and later used for normalization during inference to ensure consistent model behavior across different input distributions. Crucially, these variables are intended to be updated using non-gradient-based operations, as they are not involved in the optimization process and therefore should not participate in backpropagation. However, assigning tensors that retain gradient information—i.e., tensors with an active grad_fn—directly to these running statistics introduces unintended consequences. Specifically, PyTorch will treat these tensors as part of the computational graph, preserving their computation history for gradient tracking. This results in unnecessary memory retention, as the associated graph components are not released and continue to accumulate across training iterations. Over time, this improper management of computation graphs leads to inflated memory usage and constitutes a resource leak.

- **Mishandling training gradient:** Treating model execution during inference identically to training, particularly with respect to gradient computation, can result in inefficient memory utilization and unnecessary computational overhead. In inference scenarios, gradient information is not required, as no parameter updates are performed. However, if gradient tracking is not explicitly disabled, PyTorch will, by default, construct and store the computational graph and associated gradients. This leads to memory leaks and increased processing costs, especially in models with large architectures or when handling high-throughput data.

- **Using del Without Freeing Memory:** The assumption that calling del on variables is sufficient to free memory is a poor practice in DL frameworks like PyTorch. While del removes the reference to an object, it does not guarantee memory release if the object is part of the computation graph, which retains references to intermediate tensors. As a result, memory usage persists and accumulates, potentially causing performance degradation and out-of-memory errors. Proper memory management requires more than deletion; references in the computation graph must also be cleared to ensure efficient memory deallocation.

- **Repeated Group Creation Inside Loop:** The improper use of a loop by creating a new communication group in each iteration can contribute to memory leaks. In this context, a communication group refers to a collection of operations that coordinate across different devices, often used in distributed settings to ensure synchronization. When a new communication group is repeatedly instantiated within each iteration without proper deallocation of previous groups, memory leaks occur, as references to unused groups are retained, causing an accumulation of resources and leading to inefficiency.

- **Encoder-Decoder Inside Training Loop:** The placement of the encoder and decoder inside the training loop can exacerbate memory leakage if not managed properly. These components can retain intermediate results within the computational graph, preventing memory from being freed appropriately.

- **Tracing Inside Loop Without Cleanup:** In the context of ML, tracing refers to the process of recording the operations executed by a model during the forward and backward passes, typically for purposes such as optimization, deployment, or debugging. However, performing tracing repeatedly within a loop—without appropriate memory management—can lead to memory leaks. Each tracing operation generates a new version of the computation graph and may allocate additional memory. If this memory is not properly released, it can result in the accumulation of unused resources, leading to persistent memory usage across both RAM and GPU. This behavior poses a significant risk to application resource efficiency, particularly in long-running training or inference workflows.

Collectively, these practices impede efficient memory utilization, leading to excessive resource consumption and significant performance bottlenecks in ML applications. In conclusion, these code smells highlight the critical need for effective resource management, underscoring the importance of not solely relying on garbage collectors or memory-deallocation code but instead utilizing these tools appropriately.

Upon analyzing the collected posts related to the TensorFlow and Keras frameworks and identifying the associated code smells that contribute to resource leaks, Figure 4 presents the code smells identified in TensorFlow and Keras that contribute to resource leaks.

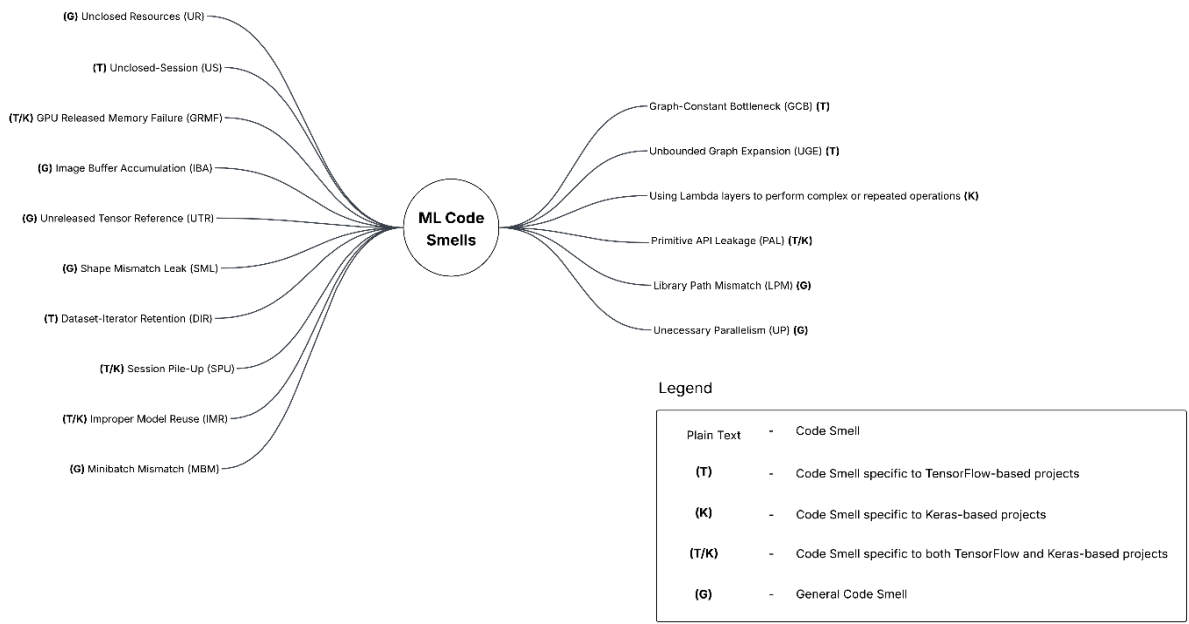

Figure 4: Code smells identified in TensorFlow and Keras

Given the close relationship between TensorFlow and Keras, and the presence of overlapping code smells in some cases, both frameworks are presented together in Figure 4. These code smells represent inefficiencies and suboptimal practices that can significantly contribute to memory leaks and result in overall inefficiencies within ML workflows. Below is a detailed analysis of the discovered TensorFlow code smells listed in Figure 4.

- **Unclosed Resource Leak:** In TensorFlow-based applications, an unclosed resource leak manifests when resources such as models, iterators, nodes, file handles, or dynamically allocated tensors are repeatedly created—especially within iterative training loops—without proper deallocation or cleanup. Each iteration may introduce new models, layers, variables, intermediate tensors, or open files without clearing the computational graph or explicitly releasing these resources, preventing the garbage collector from recovering memory. Over time, this continual accumulation leads to increased memory consumption, degraded performance, and can precipitate out-of-memory (OOM) errors or application instability. This issue is further exacerbated when developers rely solely on mechanisms like sess.close() for cleanup, which may not fully release session-associated resources. The situation becomes especially acute if combined with infinite loops, as these perpetuate the creation of unused tensors, variables, and open file descriptors without any opportunity for resource reclamation. Effective resource management thus requires explicit deallocation of unused models, iterators, tensor data, and file handles, as well as periodic graph resetting—otherwise, long-running or large-scale workflows are at risk of memory exhaustion and degraded system efficiency.
- **Primitive API Leakage:** Directly utilizing TensorFlow core operations (e.g., tf.reshape, tf.multiply) within Keras models, without appropriate encapsulation or abstraction, can result in challenges in managing the model's structure and hinder the effective use of Keras' higher-level functionalities. This practice may also limit the portability and flexibility of the model, as TensorFlow operations are not necessarily optimized for integration within the Keras framework. Furthermore, such an approach complicates debugging and maintenance by reducing code modularity, making it more difficult to manage and update the model efficiently. Another discovered bad practice is the use of a Lambda layer for complex or repeated operations in neural network models, which can lead to resource leaks if not properly managed.

- **Using Lambda layers to perform complex or repeated ops:** Lambda layers provide a concise method for defining custom operations within Keras models. However, when employed for operations that are both complex and repeated, they can introduce inefficiencies, including the retention of intermediate tensors and the computation graph, which may not be properly deallocated after the operation concludes. Such practices can result in the accumulation of memory used by these intermediate values, particularly when the operations are frequently invoked throughout the model's execution. This failure to release unused resources prevents effective garbage collection, thereby causing memory leaks.
This smell was discovered in a user-reported issue under a TensorFlow context because tf.keras is integrated within TensorFlow, leading users to attribute memory-related problems—such as those caused by Lambda layers—to TensorFlow rather than recognizing the underlying Keras-specific implementation as the root cause.
- **Session Pile-Up:** Improper reuse of model objects and the failure to clear sessions or computational graphs can give rise to substantial resource management challenges within ML workflows, particularly in frameworks such as TensorFlow. When model objects are reused without adequately clearing or resetting prior sessions or computational graphs, the system accumulates outdated resources. These retained elements, including stale models, graphs, and sessions, continue to occupy memory and computational resources, even though they are no longer necessary for ongoing operations. Over time, this can result in memory leaks, as the system does not release the memory occupied by these superfluous elements. Furthermore, neglecting to clear sessions or computational graphs exacerbates inefficient memory utilization, as these objects persist in the background, consuming both RAM and GPU memory, thereby impeding system performance.
- **Dataset-Iterator Retention:** The repeated creation of dataset iterators or operations and the subsequent feeding of model.predict with a tf.data.Dataset may result in resource leaks due to the continual creation of iterators. Specifically, when a tf.data.Dataset is passed to model.predict within a loop, a new iterator is instantiated for the dataset during each iteration. If these iterators are not explicitly managed or closed, they remain in memory, consuming resources even after they are no longer required.
- **Unreleased Tensor Reference:** Improper handling of tensors, particularly the failure to dispose of them correctly, exacerbates resource management issues. When tensors are reassigned or stored in variables without releasing the memory of the previous instances, the memory they occupy remains allocated unnecessarily. This improper use of assignment prevents the proper deallocation of memory, causing tensors to persist in memory even after they are no longer required.
- **Unclosed-Session Leak:** If sessions are not properly closed or explicitly deleted, these resources remain allocated, consuming memory even after they are no longer needed. Relying solely on garbage collection to manage unused resources can be problematic, as its timing is not immediate or guaranteed, resulting in resources persisting in memory for longer than necessary. Furthermore, the failure to properly delete models or other memory-intensive objects exacerbates this issue, contributing to memory leaks
- **Image Buffer Accumulation:** Improper resource management within the `augment_images` function can result in resource leaks when resources such as memory or computational power are not properly deallocated after use. For instance, if the function generates augmented images but neglects to release the memory allocated for these images once they are no longer required, the memory may remain occupied, gradually accumulating over time. This issue may arise if augmented images are stored in memory without proper clearing or if temporary variables holding image data are not explicitly deleted or reset. When the function is called repeatedly (e.g., within a loop for processing multiple image batches), unused resources can accumulate, eventually leading to excessive memory consumption and inefficient resource utilization. Moreover, if the images are processed on the GPU, failing to offload them to the CPU or clear GPU memory after each iteration can further exacerbate the problem, resulting in GPU memory leaks.
- **Unbounded Graph Expansion:** Reusing the default graph inefficiently and adding nodes in a loop without managing the graph can result in the continuous growth of the computational graph, as new nodes are added in each iteration without clearing or resetting the graph between loops. This accumulation of nodes, especially when the graph is not explicitly managed, leads to the retention of unnecessary operations, consuming memory and computational resources over time. As the graph grows, the system struggles to deallocate unused resources, resulting in inefficient memory utilization and potential memory leaks, which degrade system performance and increase memory consumption.
- **Graph-Constant Bottleneck:** Loading large constants directly into the graph and embedding data inefficiently exacerbates these issues. Large constants or datasets embedded directly into the computational graph may not be released after their intended use, as they are tightly integrated into the graph structure. This practice leads to the accumulation of large, unused memory objects, which are retained even after they are no longer needed for computation. As a result, these constants and data remain in memory, further increasing memory consumption and exacerbating memory leaks.

- **GPU Released Memory Failure:** Failure to manage GPU memory effectively arises when memory allocated for tensors, models, or other objects on the GPU is not explicitly released after use. In GPU-intensive tasks like deep learning, where GPU memory is limited, improper deallocation of these resources leads to memory leaks, resulting in inefficient memory utilization. Over time, unused memory accumulates, causing out-of-memory (OOM) errors, performance degradation, or even training halts. An overreliance on garbage collection worsens this issue, as it primarily targets CPU memory and is not always efficient at managing GPU memory. Relying on garbage collection to reclaim GPU memory may delay the release of unused resources, preventing timely memory cleanup. This results in further inefficiencies, as the GPU remains clogged with unnecessary tensors, slowing down computations and potentially causing resource bottlenecks in ML tasks.
- **Shape Mismatch Leak:** In DL workflows, tensors must often be reshaped to align with the specific requirements of operations such as matrix multiplications or convolutions. Failure to reshape tensors appropriately can lead to memory allocations that exceed the necessary amount, resulting in the retention of more memory than required. This inefficient memory usage occurs when large tensors are kept in memory despite the availability of smaller alternatives. Over time, such practices can accumulate, leading to memory leaks as resources remain unnecessarily allocated, consuming valuable memory space.

The following section provides a detailed analysis of the Keras-specific code smells illustrated in Figure 4. These smells are primarily observed in applications utilizing the Keras framework and reflect recurring issues in model construction, training, and inference.

- **Unclosed Resource Leak:** The failure to release resources, such as model objects, session data, or intermediate computation graphs, after each iteration of training or evaluation. This neglect leads to the accumulation of unused objects in the memory, causing memory leaks and reduced performance over time.
- **Improper Model Reuse:** The improper reuse of models, such as the generator and discriminator in a Generative Adversarial Network (GAN), without appropriate clearing or resetting, can accumulate unnecessary computation graphs, weights, and other resources. This leads to increased memory consumption and diminished computational efficiency, as redundant data occupies system resources. Similarly, the repeated recreation of layers or computation graphs in each iteration leads to excessive memory allocation, contributing to performance degradation over time. Both issues stem from inadequate resource management, which can cause memory leaks and inefficient utilization of computational resources, ultimately compromising the stability and scalability of ML workflows.
- **Using Lambda Layers to perform complex or repeated operations:** Utilizing Lambda layers for complex or repeated operations, rather than their intended purpose of executing simple, quick tasks, can lead to inefficiencies in model performance. Lambda layers are designed for lightweight, single-use functions, and repurposing them for more intricate or recurrent operations can result in unnecessary computational overhead. This misapplication can degrade model efficiency and hinder the scalability of ML systems, as Lambda layers may not be optimized for such tasks, ultimately impacting the overall execution speed and resource utilization.
- **Minibatch Mismatch:** Using excessively large minibatch sizes can significantly increase memory demand, which may compromise the efficiency and stability of ML models. While larger minibatches can accelerate training by increasing computational throughput, they also require more memory to store intermediate activations, gradients, and other tensors during forward and backward passes. This heightened memory demand can lead to out-of-memory errors, especially when working with limited hardware resources such as GPUs. Moreover, the inefficiency of managing large minibatches may hinder model scalability, as the computational overhead increases and the available memory is exhausted more quickly. Thus, selecting an appropriate minibatch size is crucial for balancing memory utilization and training performance, ensuring that memory resources are used effectively without compromising model stability.
- **Library Path Mismatch:** Improper environment configuration for CUDA paths, such as mismatched entries in LD_CONFIG_PATH and PATH, can lead to resource leaks by disrupting the proper loading and initialization of CUDA libraries and drivers. When the system fails to correctly locate or load the necessary CUDA components, it results in inconsistent GPU resource management, including improper allocation and deallocation of memory. This misconfiguration can cause GPU memory to remain allocated without being released, leading to memory leaks and inefficient GPU utilization. Additionally, repeated attempts to load incompatible or incorrect versions of CUDA libraries further exacerbate resource leakage, ultimately compromising system stability and performance.
- **Unnecessary Parallelism:** Using the parallelization setting n_jobs=-1 in GridSearchCV to parallelize model training across all available cores can contribute to resource leaks and inefficient resource utilization. By utilizing all cores simultaneously, the system may exhaust available memory and processing power, particularly in environments with limited resources or when other processes are running concurrently. This excessive parallelization can result in the accumulation of unused or improperly managed processes and threads, leading to memory leaks as the resources are not properly released once the parallel tasks are completed. The unchecked allocation of resources across multiple cores without appropriate management increases the likelihood of resource contention and can exacerbate the occurrence of memory leaks, ultimately impairing system stability and performance.

- RQ2: Under which category can the identified code smells be categorized?

The identified code smells are systematically categorized based on the underlying root causes of resource leaks. Specifically, a detailed analysis of each leak case was conducted to determine its root cause, and subsequently grouped cases exhibiting similar causes were grouped into distinct categories. This root cause-based categorization offers valuable insights by highlighting the fundamental mechanisms that lead to resource leaks. Figure 5 illustrates each identified code smell in the PyTorch-based applications, along with its associated category and sub-category.

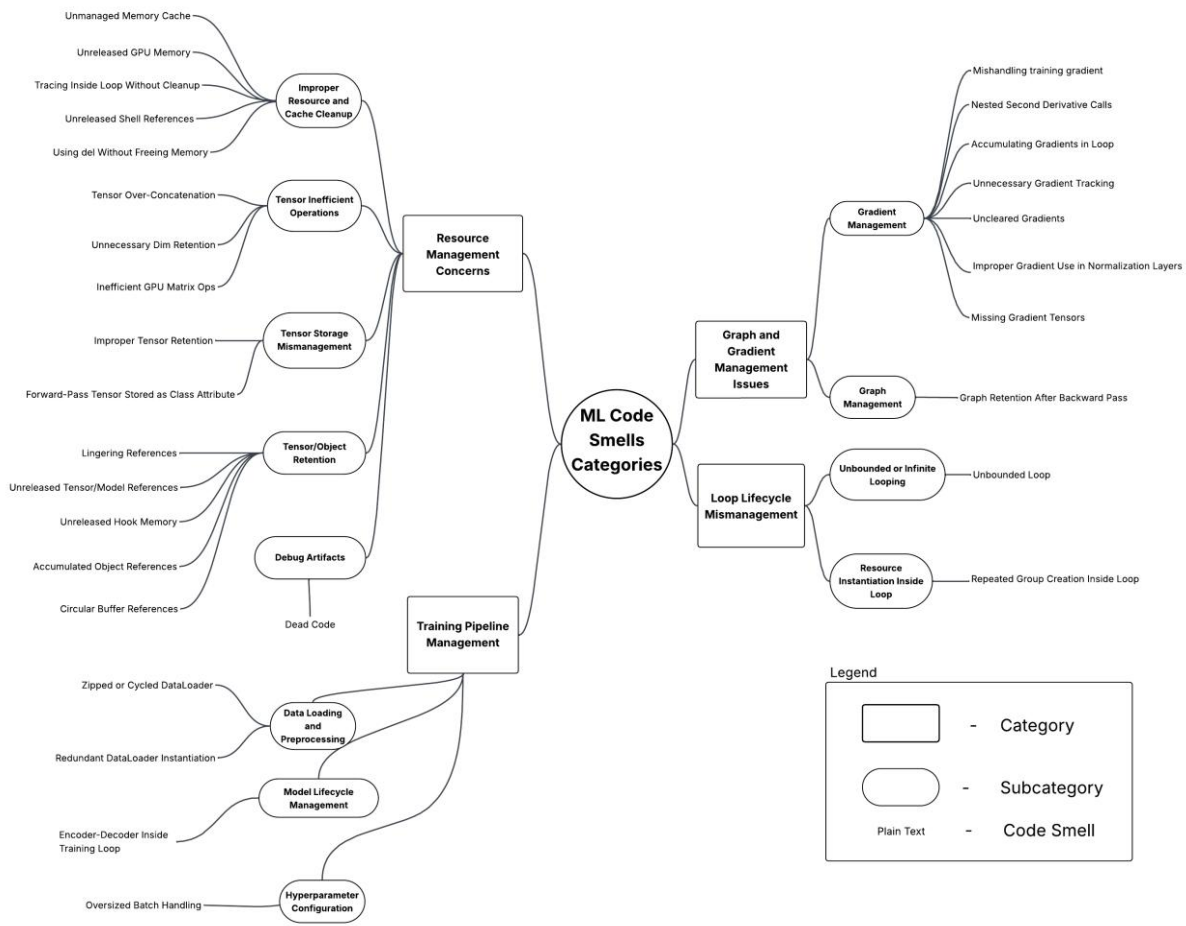

Figure 5: PyTorch code smells categories and sub-categories

The categorization of code smells in PyTorch is based on the root causes, described as follows:
- Resource Management refers to the disciplined and efficient control of computational and system resources—such as memory (CPU/GPU), tensor storage, runtime sessions, file handles, and threads—within ML applications. Code smells in this category result from improper allocation, excessive retention, or failure to release resources after use. These issues may manifest as memory bloat, redundant computation, or system-level failures such as out-of-memory errors. In ML-specific contexts, such smells often arise from mismanagement of tensors, models, intermediate computations, caches, or debugging artifacts. They typically reflect deeper violations in lifecycle management, memory discipline, or computational graph control, and can significantly compromise performance, scalability, and application stability.
- Graph and Gradient Management Issues refer to defects in the construction, usage, and lifecycle handling of computational graphs and automatic differentiation mechanisms in ML frameworks. These issues arise when gradient tracking or graph retention is either misapplied or mismanaged, leading to incorrect learning dynamics, resource leaks, or degraded performance. Problems in this category affect the correctness, efficiency, or scalability of training and inference by disrupting the intended behavior of autograd systems or by introducing unnecessary computational overhead.
- Training Pipeline Management refers to concerns related to the structural design, configuration, and orchestration of ML training workflows. This category encompasses inefficiencies, misconfigurations, and poor practices in managing the end-to-end training process, including model instantiation, data handling, training execution, and hyperparameter tuning. Smells in this category typically degrade training efficiency, lead to resource wastage, cause silent failures in learning, or produce models that generalize poorly.
- Loop Lifecycle Mismanagement refers to code behaviors where memory inefficiencies, resource leaks, or performance degradation result from improper control over the creation, execution, or termination of iterative constructs, particularly in the context of ML workflows. This category captures smells arising from incorrect loop boundaries, unregulated growth in resource allocation across iterations, or failure to release temporary structures created within loops.

Figure 6 provides an overview of the distribution and prevalence of code smell categories within the PyTorch framework.

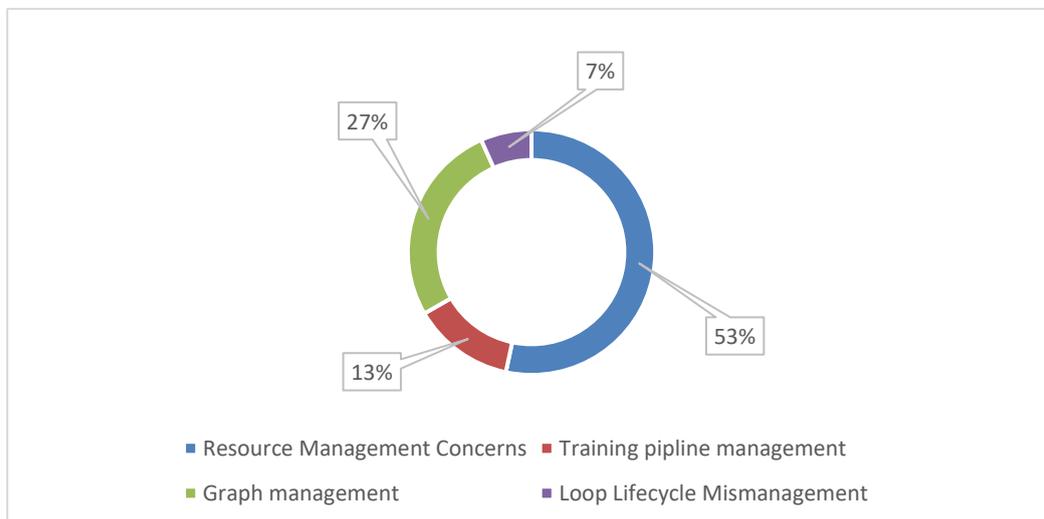

Figure 6: Distribution of Code Smell Categories Observed in PyTorch

Figure 6 presents the distribution of the thirty code smells identified as root causes of resource leaks in PyTorch-based applications. The categorization is grounded in an analysis of the underlying causes of each smell, offering an indirect but principled view of how specific practices contribute to resource inefficiencies. The results show that 53% of the smells are associated with the Resource Management Concerns category, underscoring a critical need for ML practitioners to strengthen their understanding and application of resource handling principles. The Graph and Gradient Management Issues category accounts for 27%, revealing substantial challenges in managing gradient flow and computational graph lifecycles. Training Pipeline Management contributes 13%, highlighting occasional misconfigurations in the broader training workflow. Lastly, Loop Lifecycle Mismanagement is responsible for 7% of the identified causes, indicating that while less frequent, improper loop control still plays a role in resource leaks. These findings suggest that efforts to improve resource efficiency in ML should prioritize addressing foundational issues in resource and gradient lifecycle management.

The categorization process for the TensorFlow and Keras frameworks followed a similar approach to that of PyTorch, beginning with the grouping of related code smells based on their underlying causes in ML applications, followed by the assignment of these smells to the appropriate categories. Figure 7 illustrates each identified code smell in the TensorFlow and Keras-based applications, along with its associated category and sub-category.

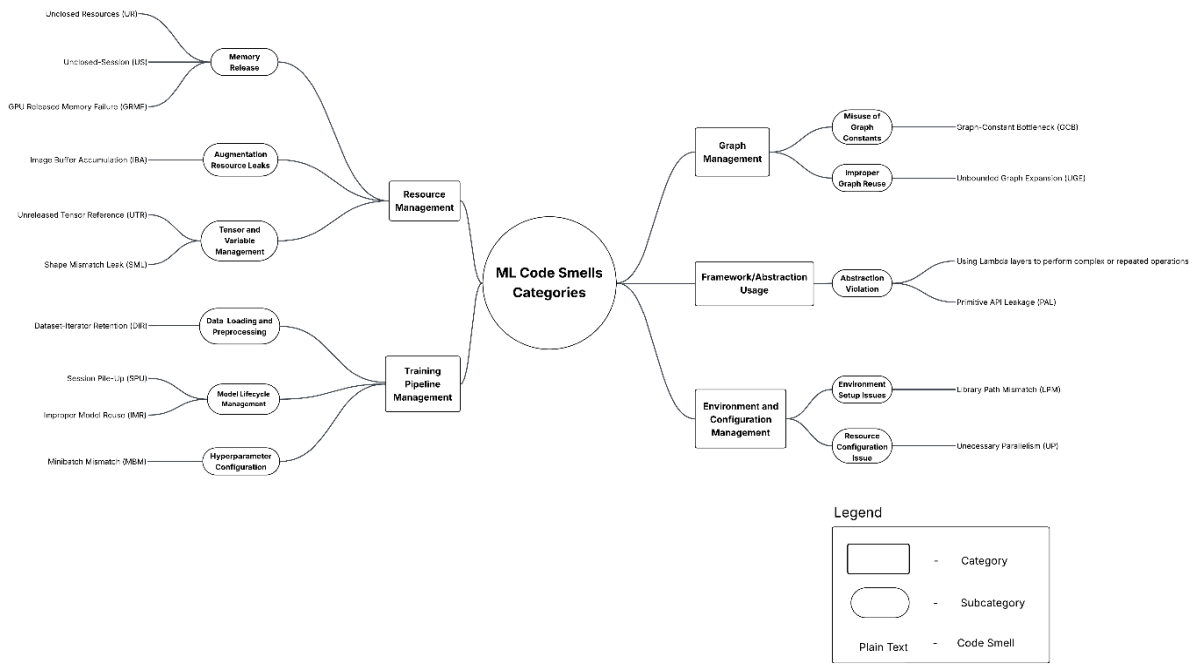

Figure 7: TensorFlow and Keras code smells categories

Smells labeled as (T/K) in Figure 7 indicate patterns that appear in both TensorFlow and Keras environments, underscoring architectural or usage commonalities between them. Notably, out of the 18 identified code smells, 5 (approximately 28%) are shared between the two frameworks. These shared smells predominantly fall under categories such as Resource Management, Framework/Abstraction Usage, and Graph Management, suggesting that improper handling of computational graphs, abstraction violations (e.g., using Lambda layers for repeated logic), and resource retention are persistent concerns irrespective of the framework's abstraction level. Furthermore, 3 out of the 6 top-level categories (50%)—namely Resource Management, Graph Management, and Framework/Abstraction Usage—contain shared smells, pointing to foundational issues in both frameworks related to memory, computation, and software architecture. This overlap reflects the fact that Keras, while abstracting TensorFlow operations, inherits many of its underlying behaviors and vulnerabilities. Therefore, identifying and addressing these shared smells can lead to broader, framework-agnostic best practices that promote efficiency, maintainability, and robustness in DL model development.

The Resource Management and Training Pipeline Management categories were previously detailed in the PyTorch section. The remaining categories specific to TensorFlow and Keras are described as follows:

- Graph Management refers to the practices involved in constructing, maintaining, and optimizing the computational graph used by the ML framework. Smells in this category arise when the computational graph is misused, overgrown, or improperly maintained, resulting in increased memory consumption, slower runtime performance, difficulty in graph serialization, or challenges in reusing models during inference or deployment.
  Mismanagement may manifest as the embedding of large constants directly into the graph, failing to separate training and inference components, continuously adding operations during iterative loops, introducing non-exportable nodes, or neglecting graph cleanup.
- Framework/Abstraction Usage refers to the appropriate utilization of high-level APIs, interfaces, and idiomatic constructs provided by ML frameworks such as TensorFlow, PyTorch, or Keras. Smells in this category arise when developers circumvent these abstractions by using low-level primitives, manual operations, or direct device control, thereby breaking the abstraction boundaries intended by the framework designers.
- Environment and Configuration Management concerns the correct setup and control of the application. Smells in this category arise when the runtime environment is improperly configured, leading to execution errors, inconsistent results, or performance bottlenecks. This includes version conflicts, incorrect environment variables, malformed configuration files, misconfigured resource limits, or problematic containerization setups.

By structuring these categories, this study provides insights into recurring inefficiencies in TensorFlow-based workflows and highlights areas where developers can improve resource utilization and model efficiency. Figure 8 illustrates the distribution and prevalence of these code smells within the TensorFlow framework categories.

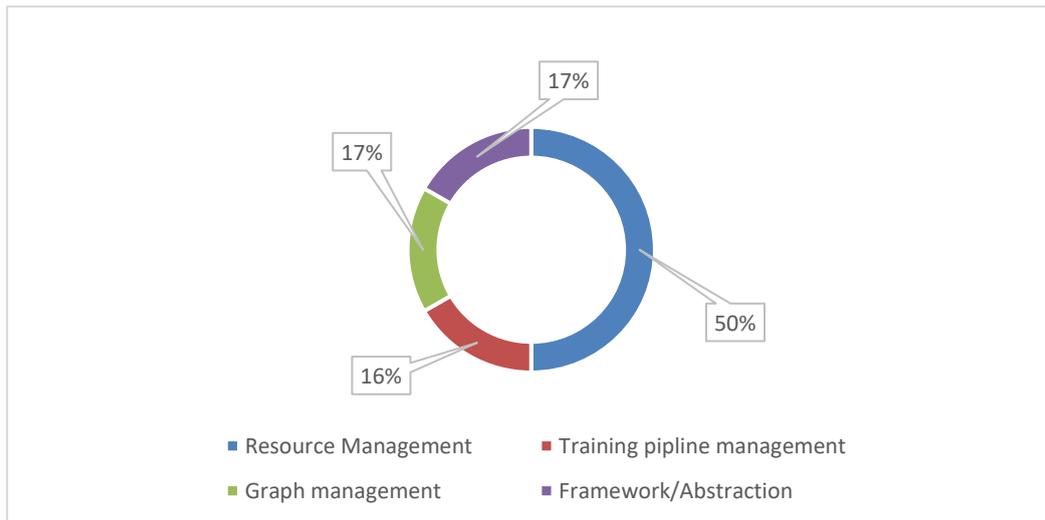

Figure 8: Distribution of Code Smell Categories Observed in TensorFlow (N=12)

The distribution of the twelve code smells contributing to resource leaks in TensorFlow highlights key areas of inefficiency within ML workflows. Resource Management (50%) emerges as the predominant factor, indicating that improper handling of computational resources, such as inefficient deallocation, is the most frequent source of leaks. Training Pipeline Management (16%) highlights issues related to data processing inefficiencies, including excessive memory retention during model training. The equal representation of Graph Management (17%) and Framework/Abstraction (17%) suggests that flaws in computational graph construction and the design of abstraction layers play a significant role in leak occurrences. This distribution underscores the need for targeted mitigation strategies, emphasizing resource management as a priority while addressing systemic inefficiencies in TensorFlow-based application development.

Figure 9 illustrates the distribution and prevalence of the identified Keras code smells within the Keras framework categories.

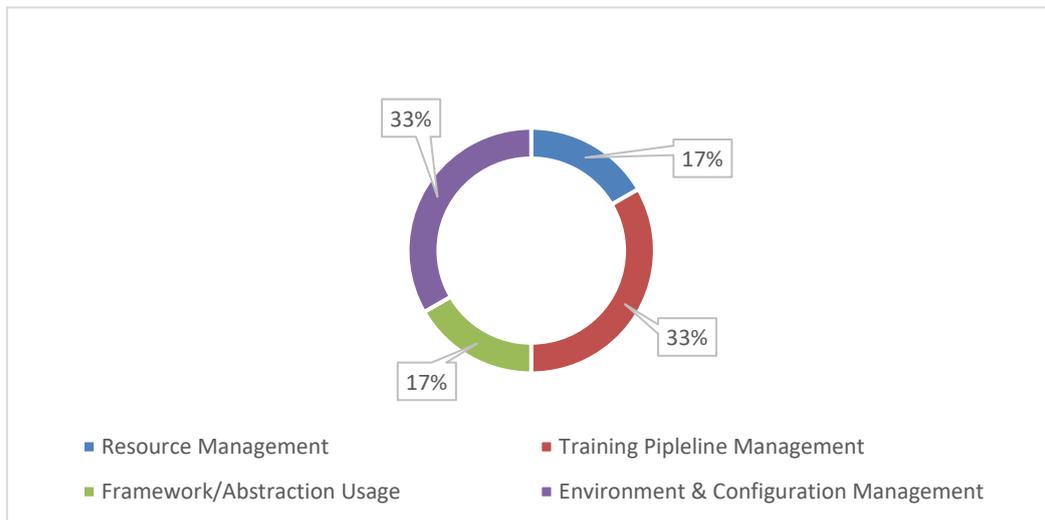

Figure 9: Distribution of Code Smell Categories Observed in Keras (N=5)

The distribution of the five code smells contributing to resource leaks in Keras reveals several critical inefficiencies within ML workflows. Training Pipeline Management (33%) constitutes a significant portion, indicating that issues such as improper data handling and memory-intensive operations during training frequently lead to resource leaks. Environment & Configuration Management (33%) highlights the role of misconfigured dependencies, version mismatches, and inefficient session management in contributing to leaks. Resource Management (17%) reflects instances where memory allocation and deallocation are mishandled, leading to persistent resource exhaustion. Similarly, Framework/Abstraction Usage (17%) signifies challenges related to improper utilization of Keras abstractions, including inefficient model instantiation and object persistence. This distribution highlights the need for targeted mitigation strategies, especially model training processes and environmental configurations.

The identified code smells can be alternatively categorized according to their applicability to general or specific ML frameworks. As illustrated in Figure 3, thirty distinct ML code smells are grouped into two categories: general smells, denoted by "G," and those specific to PyTorch, indicated as "P." General smells ("G") refer to those that may manifest across multiple ML frameworks, reflecting common inefficiencies and pitfalls inherent to ML development in general. In contrast, framework-specific smells ("P") denote issues that are uniquely associated with particular characteristics or design patterns of a given framework, in this case, PyTorch, and do not commonly appear outside this environment. Among these, twenty-one smells (70%) were classified as general, whereas the remaining nine (30%) were exclusive to PyTorch. The predominance of general smells underscores that many inefficiencies, such as memory mismanagement, resource leaks, improper tensor handling, and redundant computations, are widespread across ML codebases, regardless of the framework utilized. Conversely, PyTorch-specific smells primarily arise from characteristics unique to its dynamic computation graph and training mechanisms, including improper gradient application in normalization layers, nested second-derivative invocations, and inefficiencies related to the DataLoader. This bifurcation elucidates a dual perspective: although most code quality issues reflect universal challenges inherent to ML development, a noteworthy subset is attributable to framework-specific complexities.

Figure 4 categorizes 16 distinct ML code smells into four categories: general smells (denoted as "G"), TensorFlow-specific smells ("T"), Keras-specific smells ("K"), and those shared by both TensorFlow and Keras ("TK"). Of these, seven were categorized as general, representing approximately 44% of the total, while the remaining nine (56%) were framework-specific. Within the framework-specific group, four smells (25%) were exclusive to TensorFlow, one smell (6%) was unique to Keras, and four smells (25%) were shared between TensorFlow and Keras. General smells, such as unreleased resources, image buffer accumulation, and improper model reuse, represent inefficiencies prevalent across ML frameworks. Conversely, TensorFlow-specific smells include issues such as graph-constant bottlenecks and unbounded graph expansion, whereas Keras-specific smells are exemplified by improper session file use. The shared drawbacks of TensorFlow and Keras, such as primitive API leakage and unnecessary parallelism, highlight the challenges arising from overlapping functionalities in these frameworks. This categorization emphasizes the importance of adopting both universal best practices and framework-specific diagnostic approaches to effectively address the diverse spectrum of ML code quality issues. Many developers utilize the tf.keras API frequently observes anomalous memory growth during model training, inference, or when employing lambda layers for custom operations. This phenomenon manifests as tensors that persistently reside in memory without being appropriately released, leading to a gradual increase in resource consumption over time. Such behavior is commonly attributed to a "TensorFlow memory leak" owing to the visible escalation of memory usage within TensorFlow's runtime environment. However, a detailed analysis revealed that the underlying cause was primarily associated with Keras's Lambda layers, which retained intermediate tensors and computational graph references beyond their intended lifecycle. This retention impedes effective garbage collection, thereby contributing significantly to memory leaks within the TensorFlow execution context.

- RQ3: What are the recommended best practices (coding patterns) for preventing code smells associated with resource leaks in PyTorch, TensorFlow, and Keras-based ML applications?

This research not only sheds light on code smells contributing to resource leaks in PyTorch, TensorFlow, and Keras–based apps but also aims to provide practical solutions to address these issues. By extending the empirical analysis, the study goes beyond identifying the code smells responsible for resource leaks to uncover best practices that can help prevent these inefficiencies. The findings contribute to the field of software engineering by offering actionable strategies for developers to avoid the pitfalls leading to resource leaks. Of particular note, the PyTorch framework displayed the highest occurrence of code smells linked to resource leaks; however, it also featured the most extensive set of best practices for addressing these concerns, as identified through analysis of the collected PyTorch-related posts. Figure 10 presents recommended best practices for mitigating resource leaks in PyTorch-based applications, based on insights extracted from the analysis of collected PyTorch-related posts.

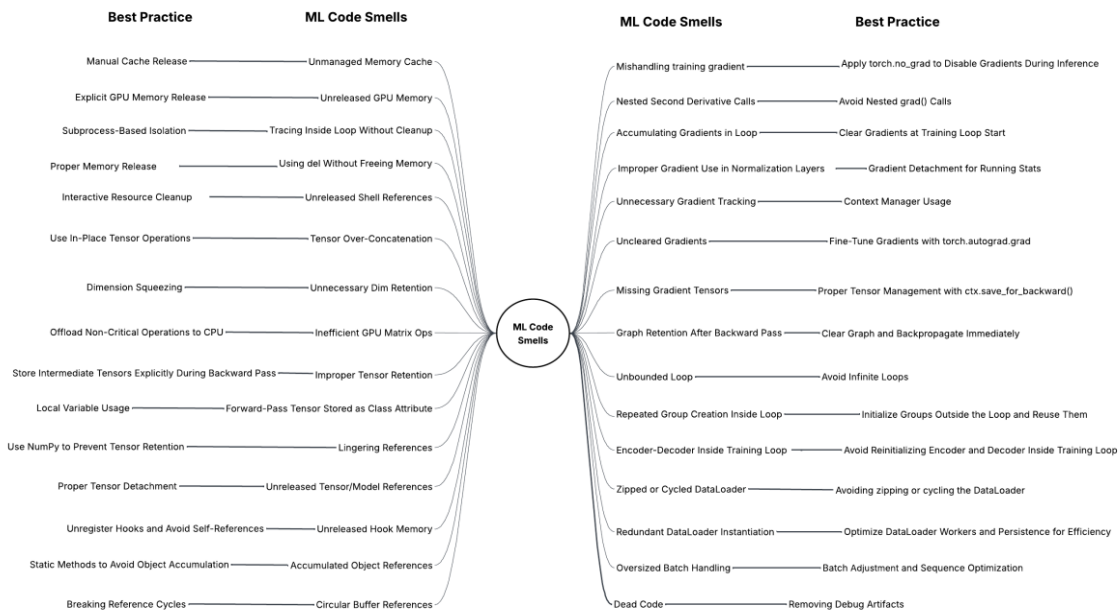

Figure 10: PyTorch best practices

To avoid memory waste and resource leaks in PyTorch-based applications, it is essential to adopt best practices that minimize code smells and ensure the efficient use of computational resources.

- **Context Manager Usage:** Using context manager statements, such as the with statement, is a best practice for managing resources efficiently in PyTorch. Context managers provide a structured approach to ensuring that resources are properly acquired and released, even in the presence of exceptions or errors. This is particularly important when addressing the excessive reliance on gradient tracking management, such as torch.no_grad(), in PyTorch. When torch.no_grad() is applied inconsistently or omitted where necessary, it can lead to unnecessary resource retention, including excessive memory usage and untracked gradients, which can negatively impact the efficiency of ML workflows. By leveraging context managers, developers can ensure that gradient tracking is correctly disabled during inference or evaluation phases and automatically restored afterward, reducing the risk of resource leaks.

  In PyTorch, automatic differentiation retains computational graphs by default to facilitate gradient-based optimization. However, during inference, gradient computation is not required, and failing to disable it leads to excessive memory retention and inefficient resource utilization. By enclosing inference operations within a torch.no_grad(): block, developers explicitly prevent the storage of intermediate tensors and computation graphs, ensuring that memory is efficiently released once the inference process is completed. This approach significantly reduces memory overhead, enhances model performance, and mitigates resource leaks, thereby improving the overall efficiency and stability of ML applications in PyTorch.

- **Unregister Hooks and Avoid Self-References:** Addressing the failure to release GPU memory after creating instances of a module with a registered forward hook is very important to avoid resource leaks. This could be accomplished by unregistering forward hooks after use and minimizing the use of self within hooks. When forward hooks are registered on model components, they create references that can persist even after the intended operation is completed. These lingering references can prevent the garbage collector from properly releasing memory, leading to GPU memory retention and potential memory leaks. By unregistering the hooks once they are no longer needed, developers ensure that these references do not persist beyond their intended scope. Additionally, minimizing the use of self within hooks prevents the creation of unintended references to the model, further optimizing memory management. Implementing this practice significantly reduces the likelihood of memory not being freed after the module and its associated hooks are no longer in use, thereby improving the overall memory efficiency and performance of PyTorch-based ML workflows.

- **Optimize DataLoader Workers and Persistence for Efficiency:** Improper management of the DataLoader with multiple workers can lead to inefficient resource utilization, particularly in terms of memory consumption and CPU overhead. A best practice to address this issue is to set persistent_workers=False in the DataLoader and limit the number of workers to zero when minimal

- parallelism is necessary. Setting persistent_workers=False ensures that worker processes are not retained across epochs, thereby preventing unnecessary memory usage from accumulating. Additionally, configuring the number of workers to 0 results in data being loaded in the main process, reducing the computational overhead associated with parallel data loading when parallelism is not essential. This practice contributes to the optimization of memory and CPU usage, promoting efficient resource allocation and minimizing inefficiencies associated with improper DataLoader management in PyTorch. A single DataLoader ensures that data is loaded in a more straightforward and controlled manner, reducing the overhead associated with managing multiple iterators or datasets. This approach minimizes the risk of data duplication, memory leaks, and inefficient resource allocation, leading to more efficient data loading and better resource management. By avoiding the use of zipping or cycling techniques, developers can enhance the overall performance and stability of their ML workflows, ensuring that the model optimally receives data.
- **Proper Tensor Detachment:** Improper handling of resources such as tensors can result in significant inefficiencies, particularly in terms of memory retention and the unnecessary buildup of computational graphs. The best practice to address these issues is to detach tensors, such as d_loss.detach() or loss.detach(), before storing them, and to ensure that hidden layers are detached when necessary. Detaching tensors prevents the retention of unnecessary parts of the computational graph that are no longer required for backpropagation. This is particularly important in scenarios where tensors are stored or manipulated outside the scope of the forward pass, as it ensures proper memory management by freeing resources once the tensor's gradients are no longer needed. Furthermore, detaching hidden layers as needed prevents them from being unintentionally retained in the graph, further optimizing memory utilization. This practice effectively reduces the risk of memory leaks and inefficiencies, supporting optimal resource management in PyTorch-based ML workflows.
- **Avoiding zipping or cycling the DataLoader:** Zipping or cycling the image DataLoader can introduce unnecessary complexities and inefficiencies in the data loading process, particularly in terms of memory usage and computational overhead. This bad practice often results in redundant data retrieval, where multiple iterators or datasets are merged or cycled, leading to increased memory consumption and slower data access. The best practice of using a single DataLoader to handle image data provides an effective solution to this issue by streamlining the data loading process.
- **Use In-Place Tensor Operations:** As the overuse or frequent concatenation of tensors along a specified dimension can lead to inefficient memory usage and potential resource leaks, the recommended best practice is to employ in-place operations or functions that avoid creating new tensors unnecessarily. In-place operations refer to functions that modify the contents of a tensor directly without allocating additional memory for a new tensor. Examples of in-place operations include torch.add_(), torch.mul_(), or tensor.resize_(), where the underscore (_) signifies that the operation is performed directly on the tensor, modifying it in place. In contrast, operations like torch.cat create new tensors to store the concatenated result, which can lead to increased memory consumption, especially when dealing with large-scale tensors. By employing in-place operations or functions that avoid the creation of new tensors, memory is used more efficiently, as no additional memory is allocated for the result.
- **Explicit GPU Memory Release:** Running a model without properly releasing GPU memory can lead to inefficient memory usage and potential memory leaks, particularly during long training or inference sessions. The best practice of periodically call torch.cuda.empty_cache() helps mitigate this issue by explicitly releasing unused GPU memory. While PyTorch's garbage collector automatically deallocates tensors when they are no longer needed, GPU memory may not be immediately reclaimed, resulting in fragmentation and suboptimal memory utilization. By invoking torch.cuda.empty_cache(), developers can manually free up unoccupied memory blocks, ensuring that previously allocated memory is returned to the GPU memory pool for reuse. This practice is especially crucial in scenarios involving large models or extensive datasets, where untracked memory consumption could lead to memory overflow errors or significant slowdowns. Regularly calling torch.cuda.empty_cache() improves memory management by reducing the risk of inefficient memory retention and memory leaks.
- **Proper Memory Release:** The %xdel magic command in Python is a crucial practice for preventing resource leaks when working with variables in an interactive environment. Unlike the standard Python shell, where variables persist in memory until the interpreter session ends or they are explicitly deleted, Python provides the %xdel command to remove variables along with all associated references. This approach ensures that memory is properly freed, reducing the risk of resource leaks caused by lingering objects that are no longer needed. When variables are not explicitly deleted, especially in long-running interactive sessions, they can accumulate and consume excessive memory, potentially leading to degraded system performance. By employing %xdel, users can proactively manage memory, preventing unintended retention of objects and improving overall resource efficiency in computational workflows.
- **Store Intermediate Tensors Explicitly During Backward Pass:** Directly saving tensors in a data

structure without utilizing proper context management can lead to resource leaks, as PyTorch's dynamic computation graph retains unnecessary references to tensors, preventing their efficient deallocation. To address this, the save_for_backward method provided by the ctx context should be used when implementing custom autograd functions. This method ensures that intermediate results are stored efficiently within the computational graph while allowing PyTorch's automatic memory management to properly track and release them when they are no longer required. By leveraging save_for_backward, developers can prevent unintended tensor retention.

- **Removing Debug Artifacts:** Leaving debugging-related code, such as torch.autograd.detect_anomaly(True), in production or training environments can lead to significant resource inefficiencies, including potential memory leaks. While anomaly detection is useful during the development and debugging phases to identify gradients or operations that might cause issues, it introduces additional overhead when enabled during training. Specifically, enabling detect_anomaly(True) forces PyTorch to track more detailed gradient computations and store additional information, which consumes extra memory and processing resources. In large-scale ML tasks, this overhead can accumulate, reducing model performance and possibly causing resource leaks due to unnecessary memory retention. By removing or commenting out debugging-related code once debugging is complete, the model can operate more efficiently, releasing memory more effectively and avoiding unnecessary computational strain.

- **Dimension Squeezing:** Unnecessary retention of dimensions occurs when tensor dimensions are preserved following a reduction operation, despite these dimensions not being required for subsequent computations. To mitigate this, it is considered best practice to manually squeeze the tensor dimensions by either using keepdim=False or explicitly calling .squeeze(). This ensures that only the necessary dimensions are preserved, thereby reducing memory consumption and improving computational efficiency. By managing tensor dimensions properly, particularly in operations involving large datasets or frequent reductions, memory leaks and performance degradation can be avoided.

- **Breaking Reference Cycles:** Circular references between buffers and objects can result in resource leaks, as the objects involved in the circular reference are not properly deallocated by the garbage collector. The best practice for addressing this issue is to use weakref.ref to create weak references to objects instead of strong references. Weak references do not increment the reference count, allowing the garbage collector to identify and free objects when they are no longer accessible. By utilizing weakref.ref, circular references are effectively broken, enabling proper memory deallocation and preventing resource leaks.

- **Static Methods to Avoid Object Accumulation:** Improperly handling accumulated references, particularly in the context of forward and backward computations, can lead to resource leaks due to the unintended retention of objects. One effective solution to this issue is to use static methods for the forward and backward computations. By defining these methods as static, they do not rely on the instance of the class, which helps avoid the accumulation of references to objects within the object's lifecycle. This ensures that intermediate results, which might otherwise persist in memory due to object instantiation, are properly handled and discarded after use. The use of static methods minimizes memory retention, allowing for better resource management and preventing memory leaks.

- **Offload Non-Critical Operations to CPU:** Performing matrix multiplication directly on the GPU without proper memory management can lead to resource leaks due to inefficient utilization of GPU memory. The best practice for mitigating this issue is to move the computation to the CPU, perform the matrix multiplication, and then transfer the results back to the GPU. This approach reduces the load on the GPU by temporarily shifting the computation to the CPU, which typically has more memory available and more efficient memory management tools. By doing so, it prevents unnecessary memory retention on the GPU, allowing for proper memory deallocation.

- **Use NumPy to Prevent Tensor Retention:** Lingering tensor references, particularly when stored in replay memory, can lead to inefficient memory usage and potential resource leaks, especially in GPU-based environments where memory resources are limited and must be carefully managed. The best practice to address this issue is to convert tensors to NumPy arrays before storing them in replay memory. This prevents tensors from maintaining strong references in memory, which may otherwise persist unintentionally. For GPU tensors, it is essential to offload them to the CPU using .cpu() before storage, ensuring that GPU memory is freed for future use. Additionally, periodically calling torch.cuda.empty_cache() helps to clear unused memory, further preventing resource leaks. By implementing these steps, memory management is optimized, preventing lingering tensor references from causing resource leaks.

- **Local Variable Usage:** Using local variables instead of class attributes to hold temporary tensors or intermediate results within the scope of a method is an effective strategy for preventing resource leaks in ML applications. Class attributes persist for the lifetime of an object, which can lead to unnecessary retention of memory if tensors or intermediate results are stored in them for temporary use. This prolonged memory retention may prevent the garbage collector from deallocating memory once the tensors are no longer required, resulting in inefficient memory utilization and potential memory leaks. In contrast, local variables are confined to the method's scope, and once the method completes, the variables are discarded, allowing the garbage collector to properly reclaim the memory. By using local variables for temporary tensors, the system avoids retaining objects longer than necessary, ensuring that memory is efficiently managed and reducing the risk of resource leaks.

- **Proper Tensor Management with ctx.save_for_backward():** In custom autograd functions within PyTorch, failing to store tensors required for gradient computation during the backward pass can lead to improper memory management, potentially resulting in resource leaks. When these tensors are not appropriately stored, they may be discarded prematurely, preventing PyTorch from tracking their memory usage and properly deallocating them. The use of ctx.save_for_backward() ensures that the necessary tensors are retained within the context, making them accessible during the backward pass for gradient calculation. This practice guarantees that memory is efficiently managed by allowing PyTorch to track and release tensors once they are no longer needed. Consequently, ctx.save_for_backward() mitigates the risk of resource leaks.

- **Manual Cache Release:** Proper management of GPU memory is critical to preventing resource leaks, particularly in DL applications where memory usage can grow significantly. One common issue arises when GPU memory is maintained as a cache without being released when no longer needed. The best practice in this case is manually releasing the GPU memory cache using torch.cuda.empty_cache(). it helps mitigate the risk of unneeded cache memory by clearing unused memory and making it available for other operations. However, it is essential to balance this with the overhead incurred when allocating new tensors, as frequent cache clearing and allocation can degrade performance. Additionally, wrapping tensors in variables only when necessary for model and loss computations ensures that unnecessary operations are not tracked, limiting memory usage and reducing the potential for resource leaks.

- **Batch adjustment and Sequence Optimization:** Improperly managing large tensors, long sequences, and large batch sizes can lead to significant resource leaks due to inefficient memory utilization. The best practice to address this issue is to reduce sequence length and batch size, which directly lowers memory consumption during training. Additionally, using mixed precision with torch.cuda.amp.autocast() can reduce memory overhead by using lower-precision data types while maintaining model performance. Ensuring that large tensors in the forward pass are detached or processed in smaller chunks further prevents the retention of intermediate results that may no longer be necessary.

- **Fine-Tune Gradients with torch.autograd.grad:** Unclearing gradients properly after multiple backward passes or creating unnecessary additional computational graphs for gradient computation can lead to significant resource inefficiencies and memory leaks. The create_graph=True argument in the backward pass creates an additional computational graph, which is useful when computing higher-order derivatives. However, this can inadvertently result in excessive memory usage when not necessary, as the graph and its associated tensors are retained in memory. A more efficient approach is to use torch.autograd.grad, which provides finer control over gradient computation. This method allows for precise gradient calculations without the need to create an extra computational graph unless explicitly required, thereby preventing the accumulation of unnecessary references and reducing memory overhead. By avoiding the creation of unneeded computational graphs and properly managing gradient computations, the practice ensures that memory is used efficiently, reducing the risk of resource leaks.

- **Clear Graph and Backpropagate Immediately:** Unnecessarily retaining the computational graph after the backward pass can lead to significant memory inefficiencies and potential resource leaks in ML models. The retention of the computation graph consumes additional memory, as it stores intermediate values required for gradient computation during backpropagation. This can be particularly problematic in scenarios involving large datasets or deep neural networks, where the graph can become exceedingly large. To mitigate this issue, it is considered best practice to avoid using retain_graph=True unless necessary. Instead, computation graphs should be cleared after use, and the backward pass should be performed immediately after computing the loss for each iteration or batch. This approach ensures that memory is efficiently managed by releasing the graph once its purpose has been served, thereby reducing memory consumption and preventing resource leaks.

- **Avoid Nested grad() Calls:** Nesting grad() calls to compute higher-order derivatives, such as the second derivative, can lead to inefficient resource management and potential memory leaks. The efficient practice involves using autograd.grad() with the create_graph=True argument, which allows for the explicit creation of the necessary graph while enabling higher-order derivatives to be computed in a more memory-efficient manner. By avoiding nested grad() calls and retaining the graph only when necessary, memory consumption is reduced, and resource leaks are prevented, ensuring more efficient use of computational resources.

- **Gradient Detachment for Running Stats:** Assigning tensors with gradient information directly to variables such as self.running_mean and self.running_covar in normalization layers can lead to unnecessary memory growth and resource leaks. These variables are typically used to store running statistics (mean and covariance) during training but should not track gradients, as they do not require any updates via backpropagation. Using.detach() on the tensors, the gradient information is effectively removed, and the computation graph is not tracked. This approach ensures that these variables do not participate in gradient computation, preventing the accumulation of memory used for unnecessary history retention. Detaching gradients in this manner promotes efficient memory usage and mitigates the risk of resource leaks.
- **Disable Gradients During Inference:** Handling the training model's gradients in the same way as during inference can lead to inefficient memory usage and unnecessary computational overhead. During inference, gradients are not required, yet by default, PyTorch will still track and store them, consuming valuable memory and processing resources. The best practice for addressing this inefficiency is to use torch.no_grad() or to set param.requires_grad = False for network parameters during inference. These methods prevent the computation and storage of gradients, optimizing both memory usage and computational efficiency. By explicitly disabling gradient tracking, resources are conserved, and unnecessary memory retention is avoided. This practice is particularly important in scenarios where inference is performed frequently or on large datasets, as it ensures that memory is not unnecessarily occupied, thus reducing the risk of resource leaks
- **Proper Memory Release:** Simply calling `del` on variables in PyTorch does not necessarily result in immediate memory deallocation, as the computation graph may still maintain references to these variables. To address this issue, it is essential to manually clear unused tensors by invoking `torch.cuda.empty_cache()`, which releases GPU memory that remains allocated due to PyTorch's caching mechanisms. Furthermore, detaching tensors from the computation graph using `.detach()` or operating within a `with torch.no_grad()` context ensures that unnecessary gradient tracking is avoided, thereby reducing memory overhead.
- **Avoid Infinite Loops:** Improper handling of an endless loop can lead to significant resource inefficiencies, particularly in iterative processes such as data loaders or model training loops. To address these issues, it is essential to avoid using itertools.cycle unless an infinite iteration is explicitly required. By adhering to this best practice, developers can ensure that iterative processes terminate as intended, allowing for effective resource management and preventing unnecessary accumulation of allocated resources.
- **Initialize Groups Outside the Loop and Reuse Them:** Improper use of loops by creating a new communication group in each iteration can lead to significant resource leaks in PyTorch-based applications. Initializing the communication group once outside the loop and reusing it as needed, developers can prevent unnecessary resource allocation and deallocation, ensuring that memory and computational resources are used efficiently.
- **Avoid Reinitializing Encoder and Decoder Inside Training Loop:** The practice of reinitializing the encoder and decoder within each iteration of the training loop can lead to unnecessary computational overhead and memory consumption, resulting in resource leaks. Instead, it's recommended to remove the encoder and decoder from inside the training loop and reuse the same instances across iterations. In this case, the unnecessary memory allocations are avoided.
- **Clear Gradients at Training Loop Start:** Failing to reset gradients at the start of each training iteration can result in the accumulation of gradients across multiple iterations, leading to potential resource leaks in PyTorch-based applications. The recommended best practice is to reset gradients at the beginning of each iteration. Thereby ensures that only the gradients relevant to the current iteration are retained.
- **Subprocess-Based Isolation:** Repeatedly performing tracing within a loop without proper memory management can lead to memory leaks due to the accumulation of unreleased resources. A recommended best practice to mitigate this issue is to perform tracing within isolated subprocesses. By encapsulating each tracing operation in a separate process, resource allocation and deallocation are confined to a controlled execution environment. This approach prevents the buildup of memory across iterations, ensures that traced models are properly discarded after use, and significantly reduces the risk of resource leaks. Subprocess-based tracing promotes better memory efficiency and improves the robustness of long-running ML workflows.

A set of best practices has been identified to address code smells in the TensorFlow framework that contribute to resource leaks. Figure 11 outlines practices aimed at mitigating these issues in TensorFlow and Keras-based ML applications.

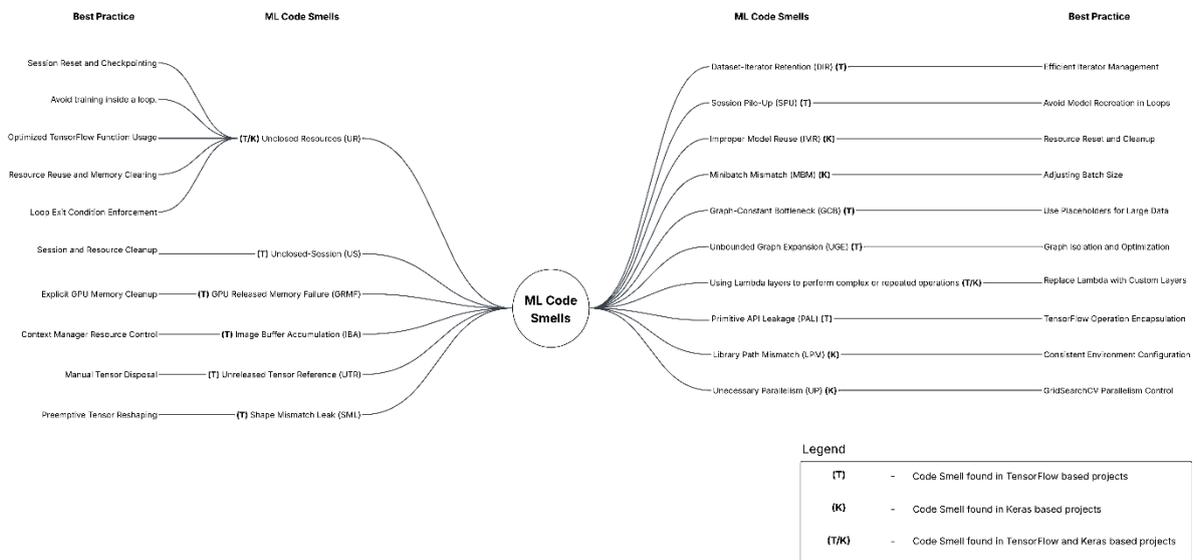

Figure 11: TensorFlow and Keras identified best practices

The practices illustrated in Figure 11 are discussed in further detail as follows:
- **Unclosed Resource smells best practices:**
  The empirical analysis of posts related to the unclosed resource smell identified four TensorFlow recommended best practices, which can be summarized as follows:
  I. **Avoid training inside a loop:** The primary best practice in ML workflows is to avoid training a model inside a loop whenever possible, as this can lead to inefficient resource utilization, including excessive memory consumption and potential resource leaks. However, when training within a loop is necessary, it is critical to apply a set of best practices to mitigate the risks of resource leakage and ensure efficient memory management. These practices include clearing sessions with K.clear_session() before instantiating the model, defining the computational graph and variables outside the loop to avoid repeated construction, and finalizing the graph using tf.Graph.finalize() to prevent further graph modifications. Furthermore, encapsulating operations within tf.function, efficiently managing data fetching, and avoiding the addition of new nodes within the loop help to optimize resource usage.
  II. **Optimized TensorFlow Function Usage:** The accumulation of uncollected objects across iterations is a detrimental practice in ML workflows, particularly during training loops in TensorFlow. To prevent such resource leaks, utilizing tf.function provides an effective solution by optimizing memory handling. By converting Python functions into TensorFlow graph operations, tf.function ensures that resources are properly managed and released after each iteration. This mechanism enables more efficient memory usage by minimizing the retention of unnecessary objects and allowing the garbage collector to effectively reclaim memory.
  III. **Loop Exit Condition Enforcement:** Ensuring that all loops have a proper exit condition is an essential best practice for mitigating resource leaks, particularly in systems handling large datasets or running over extended periods. Implementing proper loop termination ensures that resources are freed once the loop concludes, thereby preventing the accumulation of unused objects and improving system efficiency.
  IV. **Resource Reuse and Memory Clearing:** The improper handling of memory allocation, particularly in long-running loops, can lead to resource leaks, where unused tensors and computation graphs accumulate, resulting in inefficient memory utilization. The best practices outlined—reusing tensors and computation graphs, regularly using tf.keras.backend.clear_session() to free memory, enabling memory growth with tf.config.experimental.set_memory_growth(), and employing data generators or streaming pipelines—address these issues directly. Reusing tensors and graphs minimizes unnecessary memory allocations, preventing the gradual accumulation of unused resources. The use of

clear_session() ensures that unused resources are released at appropriate intervals, further reducing memory consumption. Enabling memory growth allows TensorFlow to allocate only the necessary amount of GPU memory, preventing excessive upfront memory allocation. Finally, employing data generators or streaming pipelines, only small portions of the dataset are loaded into memory at any given time, promoting efficient memory usage, particularly when working with large datasets. When combined with proper resource lifecycle management, these approaches help minimize the risk of resource leaks by ensuring that memory is allocated and deallocated in a controlled manner. As a result, they contribute to improved memory efficiency and greater stability during long-running operations.

- **TensorFlow Operation Encapsulation**: Directly using TensorFlow core operations within Keras models can lead to inefficient memory management and resource leaks, as these operations are not fully optimized for Keras' memory handling mechanisms. By encapsulating TensorFlow operations within custom Keras layers, specifically through tf.keras.layers.Layer subclasses, developers ensure that the operations integrate seamlessly with Keras' memory management system, which can track and manage memory allocation more effectively. Moreover, adopting a modular design—by breaking down the model into reusable, debuggable components—enhances the maintainability of the model and reduces the likelihood of errors that may lead to resource mismanagement. This practice not only improves the readability of the model but also supports the identification and resolution of resource leaks more efficiently. By integrating these strategies within applications, developers can significantly improve memory utilization and prevent resource leaks. This approach minimizes the risk of memory leaks by ensuring that resources are properly allocated and deallocated during the model's execution. Additionally, using consistent input shapes or padding optimizes memory usage by preventing the unnecessary creation of new nodes in the computation graph for each unique input, thus allowing for graph reuse.
- **Replace Lambda with Custom Layers**: In TensorFlow-based applications, using a Lambda layer for complex or repeated operations can lead to inefficiencies and potential resource leaks. Lambda layers often result in the creation of new, temporary operations within the computational graph, which can accumulate over time. The recommended best practice is to utilize custom Keras layers for such tasks. Developers can define explicit, reusable components that are better integrated into the TensorFlow graph, reducing unnecessary memory overhead. Custom layers allow for more efficient resource management, as they provide better control over the operations and ensure that resources are properly allocated and deallocated.
- **Avoid Model Recreation in Loops**: Improper reuse of model objects and failure to clear sessions or graphs can lead to significant resource leaks, particularly in DL frameworks. When models are repeatedly recreated within a loop without clearing the session, computational graphs and intermediate variables accumulate, leading to excessive memory usage. The best practice of calling clear_session() within the loop helps mitigate these issues by properly resetting the session and clearing the computational graph, ensuring that memory is released after each iteration.
- **Efficient Iterator Management:** Passing the entire tf.data.Dataset directly into model.predict inside a loop can lead to resource leaks due to the repeated creation of iterators for each iteration. Each time the dataset is passed to predict(), a new iterator is instantiated, and if these iterators are not explicitly managed or closed, they remain in memory. This best practice ensures that iterators are properly handled, preventing memory retention and resource accumulation, which ultimately enhances the efficiency of the inference process.
- **Manual Tensor Disposal**: Improper disposal of tensors and improper use of assignment can lead to significant resource leaks in ML workflows, as tensors may not be correctly dereferenced or deallocated before being reassigned to new operations or values. It is recommended to manually dispose of the tensors before reassigning them, as demonstrated by methods like ys.dispose() in TensorFlow.js or ensuring explicit dereferencing in Python, developers can ensure that the memory occupied by obsolete tensors is properly released. This practice allows the garbage collector to efficiently reclaim memory, thus preventing resource leaks.
- **Session and Resource Cleanup:** Ensuring proper session management and resource cleanup in TensorFlow is essential to prevent resource leaks that arise from improper session handling. The best practices to mitigate this risk include always closing sessions by calling .close() for Session or InteractiveSession after use, particularly in interactive environments like Jupyter notebooks, where memory may not be released promptly.
- **Context Manager Resource Control:** Using context managers ensures automatic management of session lifecycles, reducing the chance of resource leaks. Avoiding the passing of graphs to sessions prevents unexpected interactions and ensures the session uses the default graph. Finally, clearing sessions with tf.keras.backend.clear_session() after each session releases allocated resources, including GPU memory. These practices help prevent resource leaks, ensuring efficient memory usage and improved system performance in TensorFlow workflows. For instance, when working with file I/O or external connections, utilizing context managers (e.g., with open()) automatically handles resource cleanup, releasing resources as soon as they are no longer required. If context managers are not applicable, developers must manually dispose of resources, such as closing files or clearing memory, at the end of the task. This approach ensures that resources are freed up promptly, preventing unnecessary accumulation and memory leaks.

- **Graph Isolation and Optimization:** Reusing the default graph inefficiently, particularly by adding nodes in a loop without proper management, can lead to significant resource leaks in TensorFlow workflows. As nodes are continually added to the default graph during each iteration, the graph grows indefinitely, consuming increasing amounts of memory and computational resources. This uncontrolled expansion prevents the proper release of resources, as the default graph retains all previously added nodes, leading to memory exhaustion and degraded system performance over time. The best practice of using separate graphs and optimizing graph construction helps mitigate these issues. By creating a new graph for each function call with tf.Graph().as_default() and reusing a single graph outside of loops, the system ensures that resources are managed efficiently. Furthermore, using placeholders for variables allows for dynamic value passing without the need to repeatedly construct new nodes, promoting better resource utilization and preventing the accumulation of unused resources in the graph.

- **Use Placeholders for Large Data:** Loading large constants directly into the graph is a detrimental practice in ML workflows, as it can lead to excessive memory consumption and hinder the efficient use of computational resources. Large datasets, such as saved embeddings, often contain substantial amounts of data that, when loaded directly into the computation graph, can overwhelm available memory and prevent effective resource management. To mitigate this issue, utilizing placeholders provides an optimal solution. Placeholders in TensorFlow are special variables that allow for the deferred feeding of data into the graph during execution, rather than storing large datasets within the graph itself. This approach enables the efficient loading and processing of data in memory as needed, reducing the risk of memory exhaustion. By using placeholders, memory is not prematurely consumed by large constants, and resources can be allocated dynamically, thus preventing resource leaks.

- **Explicit GPU Memory Cleanup:** Failure to properly manage GPU memory and an overreliance on garbage collection can lead to resource leaks, particularly in ML workflows where large models and datasets are used. Without explicit memory management, objects such as models, variables, and intermediate results may persist in memory longer than necessary, consuming valuable computational resources. This can result in excessive GPU memory usage, performance degradation, and eventually memory exhaustion. To address these issues, a combination of best practices is recommended. First, manually deleting the model object with del model ensures that the model is explicitly removed from memory when no longer needed. Following this, calling gc.collect() forces Python's garbage collector to clean up any remaining objects that may not have been automatically deallocated, further reducing memory usage. Additionally, resetting the Keras/TensorFlow session using K.clear_session() explicitly frees GPU memory that may not be released otherwise, improving resource management. These practices collectively minimize reliance on automatic garbage collection, enhance memory efficiency, and prevent the accumulation of unneeded resources, thereby mitigating the risk of resource leaks and ensuring optimal performance in TensorFlow-based ML workflows.

- **Preemptive Tensor Reshaping:** Excessive memory consumption during tensor operations can arise when tensors are not reshaped appropriately, leading to unnecessary broadcasting. Broadcasting occurs when tensors of incompatible shapes are automatically expanded to match each other, which can introduce significant overhead in memory usage. This inefficiency is particularly problematic in ML workflows, where large datasets and complex models can exacerbate memory issues. Employing the best practice of using .reshape() to explicitly match tensor shapes before performing operations, unnecessary broadcasting is avoided, thereby optimizing memory utilization. This approach ensures that tensors are of compatible shapes, minimizing the need for automatic broadcasting and the associated memory overhead. As a result, it helps prevent the accumulation of unused memory and resource leaks.

The recommended best practices that can be used to replace the code smells and avoid resource leaks in Keras-based ML applications can be detailed as follows.

- **Session Reset and Checkpointing:** This best practice specifically addresses the Unclosed Resource smell from the perspective of Keras. Two effective strategies can be employed to mitigate the issue of failing to release resources after each training iteration. First, invoking keras.backend.clear_session() at the end of each training loop facilitates the release of memory by clearing the Keras session. This operation deallocates resources associated with the model, including weights, layers, and optimizer states, thereby ensuring that unused objects are removed from memory and preventing their unnecessary accumulation. Second, the use of a model checkpoint callback enables the periodic saving of model weights—typically after each epoch—allowing the training state to be preserved without retaining the full model in memory. This practice helps to optimize memory usage throughout the training process and reduces the likelihood of memory overload, contributing to more stable and resource-efficient model training.
- **Resource Reset and Cleanup:** Reusing models without proper clearing or resetting can lead to significant resource leaks in Keras workflows, as uncollected objects, such as model weights and intermediate variables, accumulate over time. The recommended best practices include utilizing tf.keras.backend.clear_session() after each iteration. This function clears the Keras session, effectively freeing memory and allowing for efficient reuse of resources in subsequent iterations. Additionally, leveraging model.fit() for training provides a structured approach to memory management by ensuring that resources are allocated and deallocated appropriately during the training process. Also, it's recommended to reuse computation graphs across iterations, instead of constructing new ones, which further optimizes memory usage and prevents the creation of redundant operations that consume unnecessary resources. It's strongly recommended to avoid dynamic operations inside loops, unless encapsulated properly, to prevent the continuous allocation of memory during each iteration. The use of stateless loss and metric functions mitigates the retention of unnecessary state, ensuring that only relevant information is maintained throughout the training process. Finally, enabling eager execution during debugging or small-scale tasks minimizes graph-related complications, allowing for more transparent and efficient memory handling.
- **Encapsulate custom operations in subclasses**: Utilizing Lambda layers for complex operations in Keras-based applications can lead to inefficient memory management and potential resource leaks. To prevent such issues, the best practice is to encapsulate complex operations in custom subclasses of tf.keras.layers.Layer. By using custom layer classes for non-trivial operations, TensorFlow and Keras are able to manage resources more efficiently, ensuring that memory is properly allocated and released as needed. Custom layers allow for better control over resource handling, enabling explicit cleanup and reducing the risk of uncollected objects persisting across iterations.
- **Adjusting Batch Size:** Inefficient minibatch sizing in Keras-based applications can lead to excessive memory consumption, particularly when working with large models or limited resources. The best practice of reducing batch size helps mitigate these issues by ensuring that memory consumption is more evenly distributed across iterations, allowing for better control over resource usage. Smaller batch sizes reduce the memory load during training, thereby minimizing the risk of memory exhaustion and facilitating proper resource cleanup after each iteration
- **Consistent Environment Configuration:** Improper environment configuration, particularly concerning CUDA paths, can lead to various issues in ML workflows, including resource leaks. Ensuring that environment variables are consistently configured and using tools like virtualenv or Docker to isolate dependencies are recommended best practices. These measures ensure that the correct libraries are used, facilitating efficient resource management and preventing the accumulation of uncollected resources.
- **GridSearchCV Parallelism Control:** The bad practice of setting n_jobs=-1 in GridSearchCV to parallelize model training across all available cores can lead to significant resource leaks and inefficient resource utilization. The recommended best practices—setting n_jobs=1 for single-threaded execution and considering alternative parallelization approaches such as asynchronous training techniques or libraries like Dask, Ray, or Python's multiprocessing—offer a more controlled approach to resource management. By limiting the number of threads or distributing the load more effectively across multiple processes or machines, these practices prevent the excessive accumulation of unused resources and allow for better cleanup, reducing the risk of memory leaks.

## 5. Discussion

To the best of our knowledge, this is the first study to focus on code smells that lead to resource leaks in ML applications. The existing literature on resource leaks does not adequately cover ML applications and their libraries, leaving a significant gap in understanding the specific challenges and intricacies involved. This lack of coverage leads to a limited comprehension of how resource leaks manifest and persist in ML environments.

This research aimed to address code smells that lead to resource leaks within ML applications developed using PyTorch, TensorFlow, and Keras, intending to identify and resolve resource leakage issues specific to these widely utilized frameworks.

By examining all relevant challenges across the three frameworks, the study methodology ensured a comprehensive analysis and provided practical recommendations to enhance resource management and mitigate leakage issues in PyTorch, TensorFlow, and Keras workflows. Furthermore, this research contributes to the broader field of software engineering by uncovering and categorizing the code smells that contribute to these leaks and discovering best practices to address the code smells that result in resource leaks. These contributions aim to improve resource efficiency, coding quality, and sustainability in ML development practices.

PyTorch exhibits a higher frequency of posts related to resource leaks compared to TensorFlow and Keras. However, this observation does not necessarily imply that PyTorch is inherently more susceptible to such issues. Interpreting this trend requires further investigation and consideration of multiple contextual factors. For example, the degree of integration with external ML and data processing libraries can influence the occurrence of resource leaks. PyTorch's widespread use in research and its compatibility with low-level libraries such as NumPy and SciPy may introduce additional complexity in resource management, thereby increasing the likelihood of mismanagement. In contrast, Keras, with its focus on high-level abstraction and ease of use, typically involves fewer integrations, potentially reducing exposure to resource-related issues. Moreover, PyTorch's close interaction with GPU resources can make the effects of resource leaks more visible, while in TensorFlow or Keras, similar issues may be less apparent due to differences in execution models and memory handling strategies.

The empirical analysis showed that 30% (9 out of 30) of the identified smells from PyTorch-related posts are specific to the framework, reflecting concerns that arise from its architecture and API design. The remaining 70% were categorized as general smells that are likely to occur in other ML frameworks as well, including TensorFlow and Keras. This distribution suggests that the presence of resource leak-related issues in PyTorch-based applications does not necessarily reflect deficiencies unique to PyTorch itself but rather highlights the general nature of many resource management problems across frameworks. Nevertheless, the comparatively larger number of PyTorch-related posts in the dataset warrants further investigation to determine whether this is due to broader adoption, higher reporting activity, or other factors influencing visibility. In conclusion, this study provides a foundation for future empirical research to determine whether PyTorch is truly more susceptible to resource leaks or if the observed patterns are mainly due to higher engagement and reporting in PyTorch-based development contexts.

To address the second research question, the categorization process was conducted in two phases. Initially, three authors independently classified the code smells into distinct categories. Following this, a discussion session was held to assess the level of agreement among their categorizations and to refine the category names. In some instances, naming was relatively straightforward, as with "Training Pipeline Management," where it was clear that the training is managed in a way that contributes to resource leaks, such as cycling the DataLoader. The first discussion did not lead to immediate consensus. After taking notes from the session, each author re-applied the categorization process independently. A second discussion session was then held to review the revised categorizations and evaluate the level of agreement. Once consensus was reached on both the categorization and category names, the process was repeated for the categorization of code smells in a new ML framework.

The categorization process is grounded in the underlying cause or type of issue that each code smell represents. Specifically, the categorization reflects a logical grouping based on the nature of the problem, facilitating the identification of root causes and potential solutions. Grouping the smells based on the type of issue allowed us to establish criteria for each category, simplifying the addition of new cases to the appropriate category. These criteria are elaborated in the response to the second research question. However, this approach occasionally posed a challenge, as some code smells may fit into multiple categories. For example, the code smell "Tracing Inside Loop Without Cleanup" could be categorized under both Resource Management Concerns and Loop Lifecycle Mismanagement. After extensive discussion, the authors decided to assign it to the Resource Management Concerns. This decision was based on their analysis, which determined that while the loop in this case refers to the training process, the training itself does not inherently lead to resource leaks. Rather, the core issue stemmed from poor resource management.

Our empirical analysis focused on uncovering the root causes of resource leaks, extending beyond surface-level keyword identification to examine the fundamental mechanisms behind these issues. For example, the repeated creation of dataset operations during inference should not be categorized as Loop Lifecycle Mismanagement, as the problem does not stem from the loop structure itself but from the mismanagement of resources within each iteration. This behavior is more appropriately classified under Training Pipeline Management, as it reflects inefficient handling of dataset operations during prediction. Repeatedly instantiating dataset iterators introduces redundant nodes into the computational graph, which—if not properly managed—can lead to memory accumulation. Over time, this buildup contributes to resource leaks and degrades system performance, underscoring the importance of robust resource management practices during the inference phase.

Our empirical study revealed that ML developers often rely on GPU-utilization metrics to detect resource leaks. However, this approach frequently causes confusion between resource leaks and optimization issues, exposing key gaps in current practices. This raises an important question: How reliable is GPU utilization as an indicator of resource leakage? These findings highlight the need for clearer guidelines to distinguish between these two issues in ML frameworks. Addressing these questions challenges the prevailing assumptions within the developer community and opens the door to a new research direction aimed at enhancing the accuracy and effectiveness of resource management in ML applications. This line of inquiry is particularly significant given the increasing complexity of ML models and the growing reliance on GPU resources, where misinterpretations can lead to inefficient resource usage and potentially impact system reliability.

This research will be expanded by examining additional ML libraries, such as JAX and ONNX. The goal is to address all aspects of code smells that lead to resource leaks in ML applications. Additionally, it provides a thorough understanding of resource leaks and suggests strategies for reducing these issues.

## 6. Validation

Validating empirical findings in software engineering is essential, particularly when investigating critical concerns such as resource leaks in ML applications developed using PyTorch, TensorFlow, and Keras. Validation strengthens the reliability and generalizability of the identified resource leak issues, the associated code smells, and the recommended best practices. Through rigorous validation, this research aims to provide practitioners with actionable, evidence-based guidance for mitigating resource leaks in real-world scenarios. This process not only reinforces the practical relevance of the findings but also enhances confidence in their applicability, ultimately contributing to the development of more efficient and robust ML applications.

The validation of this research was conducted in three phases to ensure the accuracy and reliability of the findings. In the first phase, three authors independently extracted code smells from the posts, applying the research methodology and analysis. This was followed by a discussion between the authors to verify that the identified smells were indeed those leading to resource leaks. A second discussion session was then held to finalize the validation, as full agreement was not reached initially. The second validation phase began after the code smells had been validated, with each author categorizing them independently. A first discussion session was held to present and discuss the results, but, as in the first phase, full agreement was not achieved. A second discussion session was then conducted to reach a final consensus. The final decision in both phases was made by the first author.

The final validation phase focused on confirming that the discovered best practices were appropriately linked to the identified bad practices. In this phase, the three authors independently extracted the best practices, followed by two discussion sessions to resolve any disagreements. As with the previous phases, the full agreement was not initially reached, and a second discussion session was necessary to reach a final consensus. The final decision was once again made by the first author. This validation process was carried out carefully and systematically to ensure that the process was executed rigorously and that the results were validated appropriately.

To assess the reliability of the code smell categorization, we computed the inter-rater agreement among the three authors. Considering the full set of 46 identified smells (comprising 30 discovered in PyTorch discussions and 16 smells discovered in TensorFlow/Keras discussions), disagreement occurred in four cases, resulting in a percent agreement of approximately 91.3%. To further account for chance agreement, Cohen's kappa coefficient ($\kappa$) was calculated, yielding a value of $\kappa = 0.88$. According to the commonly used interpretation scale by Landis and Koch (Landis and Koch 1977), this value indicates an almost perfect agreement between the raters. This high level of concordance supports the consistency and reliability of manual classification.

## 7. Threats to Validity

This study is subject to several potential threats to validity, which we address using the following standard classifications:

1. Construct validity concerns whether the study accurately captures the concept of resource leakage. To mitigate this, we used a clear definition—persistent resource consumption without proper release—and selected only developer discussions that included both reported leaks and their corresponding resolutions. Although keyword filtering was employed to identify relevant posts, all entries were manually validated to exclude misclassified cases (e.g., confusion with LeakyReLU). Nonetheless, some degree of interpretation bias may remain because of the informal nature of developer-reported issues.
2. Internal validity relates to the rigor of the analysis and whether uncontrolled variables may have influenced the results of the study. To reduce this threat, three authors independently analyzed and categorized the data, resolving discrepancies through discussion. Manual filtering and deduplication were also applied to ensure the integrity of the dataset. However, complex cases involving multiple interacting factors may not have been fully resolved.
3. External validity addresses the generalizability of findings. Our dataset includes real-world cases from three widely adopted ML frameworks (PyTorch, TensorFlow, and Keras) over an eight-year period. However, the findings may not generalize to less common frameworks (e.g., JAX), other programming languages, or enterprise environments not represented in public developer forums. It is also important to note that developer discussions were collected from different community platforms: PyTorch posts were sourced from the official PyTorch forum, while TensorFlow posts were sourced from both the official TensorFlow forum and Stack Overflow. Keras posts were obtained from the Stack Overflow website. This variation reflects the availability of relevant data rather than methodological preferences. The study initially focused on PyTorch, which has a centralized and active official forum. When the scope was later extended to TensorFlow and Keras, resource leak discussions were found to be sparse on their official platforms, prompting the use of Stack Overflow, which provided more substantial and relevant content than the official platforms. Although this introduces variability in the data sources, we mitigated the impact by focusing solely on technically detailed discussions of resource leak issues, regardless of the platform. Moreover, it is important to acknowledge that our findings do not encompass all possible code smells that contribute to resource leaks in ML applications using PyTorch, TensorFlow, or Keras. The scope of this study is limited to the specific sample analyzed and may not capture all potential patterns or issues that lead to resource inefficiencies in the broader ML landscape. Additionally, the best practices identified in this study were derived from this sample and may not fully reflect the diversity of coding practices across ML projects. Therefore, while the recommendations are valuable, they should be interpreted with an understanding of their dataset-specific context and limited generalizability to other populations.
4. Conclusion validity concerns the soundness of the interpretation. We grounded all identified smells and best practices in developer-reported experiences and supported them with detailed examples and consistent categorization. Although qualitative thematic analysis provides meaningful insights, it is inherently subjective. Cross-validation among multiple researchers helped reduce this threat, but some interpretive bias may remain.

## 5. Conclusion

While the field of ML has largely focused on improving model performance through metrics such as accuracy, precision, and recall, real-world deployment also demands attention to the efficiency and sustainability of ML applications. In particular, improper resource handling, such as allocating memory or computing resources without proper deallocation, can lead to resource leaks, degrade system performance, and increase operational costs. This study addresses this often-overlooked dimension by analyzing bad coding practices to identify code smells that contribute to resource leaks in ML applications developed with PyTorch, TensorFlow, and Keras.

To systematically investigate this issue, an empirical study was conducted to answer three key research questions:
- Identification of Code Smells: The first research question focused on identifying coding practices that lead to resource leaks. Through manual analysis of real-world developer discussions, 46 distinct code smells were identified across the three frameworks.
- Categorization of Smells: In response to the second research question, these code smells were categorized based on the root causes of resource leaks. Additionally, the smells were categorized according to their scope—whether they are general to ML or specific to a particular framework, highlighting the breadth and framework-dependence of resource management challenges.
- Inter-rater agreement was assessed among the three authors to ensure the reliability of the categorization process. The evaluation covered all 46 identified smells, with disagreement occurring in only four cases, resulting in a percent agreement of approximately 91.3%. To account for chance agreement, Cohen's kappa coefficient ($\kappa$) was also calculated, yielding a value of $\kappa = 0.88$. This high level of agreement supports the consistency of the classification and reinforces the validity of the categorization scheme.
- Best Practices for Mitigation: The third research question aimed to uncover corrective strategies. For each code smell, at least one best practice was identified, culminating in a total of 50 recommended practices designed to mitigate resource leaks and improve the sustainability of ML applications.

This study employed a rigorous three-phase validation process—including independent extraction, categorization, and iterative consensus-building discussions—to ensure the accuracy and reliability of the findings related to resource leaks, code smells, and best practices in ML applications developed with PyTorch, TensorFlow, and Keras. This methodological approach reinforces the practical relevance and trustworthiness of the results.

Addressing a notable gap in the literature, this work represents the first systematic empirical investigation into code smells that contribute to resource leaks in widely used ML frameworks. Prior to this research, resource management issues in ML development had received limited attention from a software engineering perspective. By identifying 46 distinct resource-related code smells and proposing 50 actionable best practices, this study offers a significant contribution to the intersection of ML and software engineering.

Beyond revealing specific anti-patterns, the study provides concrete, framework-informed guidelines that developers can adopt to improve the efficiency and long-term sustainability of ML applications. These findings not only enhance current development practices but also lay a foundation for future research on automated detection and remediation of resource-related issues in ML codebases.

**Funding:** This research did not receive any specific grant from funding agencies in the public, commercial, or not-for-profit sectors.

**Author Contributions:** The study was conceived and designed by Bashar Abdallah. Data collection was carried out by Bashar Abdallah and Gustavo Santos. Data analysis was performed by Bashar Abdallah, Martyna E. Wojciechowska, Edmand Yu, and Gustavo Santos. The manuscript was drafted by Bashar Abdallah, and all authors contributed to reviewing, editing, and approving the final version of the manuscript.

**Competing Interests:** The authors declare that there are no conflicts of interest.